\documentclass[]{JHEP3} 

\usepackage{epsfig,multicol,bbm}
\usepackage{amsmath}
\usepackage{amstext,amssymb,amsfonts, graphicx, subfigure}
\usepackage{slashed}

\newcommand\fverb{\setbox\fverbbox=\hbox\bgroup\verb}
\newcommand\fverbdo{\egroup\medskip\noindent
            \fbox{\unhbox\fverbbox}\ }
\newcommand\fverbit{\egroup\item[\fbox{\unhbox\fverbbox}]}
\newbox\fverbbox

\def\t{\tau}
\def\e{\,{\rm e}}

\def\be{\begin{equation}}
\def\ee{\end{equation}}
\def\ba{\begin{eqnarray}}
\def\ea{\end{eqnarray}}

\newcommand{\eg}{{\it e.g.~}}
\newcommand{\ie}{{\it i.e.~}}
\newcommand{\del}{\partial}

\newcommand{\ra}{\rangle}

\newcommand{\wt}{\widetilde}

\newcommand{\nn}{{\nonumber}}

\def\a  {\alpha}                
               
\def\e  {\epsilon}     \def\ve {\varepsilon}   
\def\l  {\lambda}             \def\m  {\mu}
\def\n  {\nu}          \def\s  {\sigma}        
\def\t  {\tau}

 \newcommand{\call}{\mbox{${\cal L}$}}
 \newcommand{\caln}{\mbox{${\cal N}$}}
\newcommand{\calo}{\mbox{${\cal O}$}}

\def\IR{{\hbox{{\rm I}\kern-.2em\hbox{\rm R}}}}
\def\IB{{\hbox{{\rm I}\kern-.2em\hbox{\rm B}}}}
\def\IN{{\hbox{{\rm I}\kern-.2em\hbox{\rm N}}}}
\def\IC{\,\,{\hbox{{\rm I}\kern-.59em\hbox{\bf C}}}}
\def\IZ{{\hbox{{\rm Z}\kern-.4em\hbox{\rm Z}}}}
\def\IP{{\hbox{{\rm I}\kern-.2em\hbox{\rm P}}}}
\def\IH{{\hbox{{\rm I}\kern-.4em\hbox{\rm H}}}}
\def\ID{{\hbox{{\rm I}\kern-.2em\hbox{\rm D}}}}

\def\ra{\rightarrow}

\def\del{\partial}

\newcommand{\brac}[1]{\langle #1 \rangle}

\def\tr{{\rm tr}\,}

\def\nn{\nonumber}
\def\ea{{\it et al}. }

\newcommand{\tw}{{{\widetilde w}}}

\newcommand{\trho}{{{\widetilde \rho}}}

\newcommand{\tL}{{{\widetilde L}}}
\newcommand{\tc}{{{\widetilde c}}}
\newcommand{\ttau}{{{\widetilde \tau}}}
\newcommand{\te}{{{\widetilde \ve}}}
\newcommand{\tm}{{{\widetilde m}}}

\newcommand{\nB}{{{{B}}}}
\def\tr{{\tilde r} }

\title{Non-equilibrium physics at a holographic chiral phase transition \\
 }

\author{Nick Evans$^a$, Tigran Kalaydzhyan$^b$, Keun-young Kim$^a$, and Ingo Kirsch$^b$\\
$^a$School of Physics \& Astronomy, University of Southampton,
Southampton, UK.\\ $\left. \right.$ Kavli Institute for
Theoretical Physics China, CAS, Beijing 100190, China. \\
$^b$DESY Hamburg, Theory Group,
$\ \,$Notkestrasse 85, D-22607 Hamburg, Germany.\\
    E-mail: \email{evans@soton.ac.uk}, \hspace{0.2cm}\email{tigran.kalaydzhyan@desy.de}, \\
    \hspace{1.3cm}\email{k.kim@soton.ac.uk}, \hspace{0.05cm} \email{ingo.kirsch@desy.de}}

\preprint{
           DESY-10-196\\
           SHEP-10-38 }  

\abstract{The D3/D7 system holographically describes an ${\cal
N}$=2 gauge theory which spontaneously breaks a chiral symmetry by
the formation of a quark condensate in the presence of a magnetic
field. At finite temperature it displays a first order phase
transition. We study out of equilibrium dynamics associated with
this transition by placing probe D7 branes in a geometry
describing a boost-invariant expanding or contracting plasma. We
use an adiabatic approximation to track the evolution of the
quark condensate in a heated system and reproduce the phase
structure expected from equilibrium dynamics. We then study
solutions of the full partial differential equation that describes
the evolution of out of equilibrium configurations to provide a
complete description of the phase transition including describing
aspects of bubble formation.}

\keywords{AdS/CFT Correspondence, chiral phase transition, non-equilibrium physics}

\begin{document}


\section{Introduction}

Thermal phase transitions are a crucial aspect of the evolution of
the Universe after the Big Bang and also in the physics of heavy
ion collisions. We have traditionally lacked tools to study these
transitions though in strongly coupled systems such as QCD. The
AdS/CFT correspondence \cite{Malda}, which gives a weakly coupled
string/gravity description of a class of strongly coupled gauge
theories, offers the chance to study similar transitions in
detail.

In this paper we study a simple dual of a theory with gauge fields and quarks
which has (in the presence of a magnetic field) a first order chiral symmetry restoring
phase transition as it is heated. Previous analysis of this gauge theory (using probe branes
in an AdS-Schwarzschild black hole geometry) has explored the first order
transition for equilibrium (time independent) configurations.
The heating or cooling of the system can be studied thanks to
the boost-invariant expanding or contracting plasma geometry of Janik and Peschanski \cite{Janik:2005zt}. That geometry,
which we will review below, has a moving black hole horizon describing the changing
temperature in the gauge theory. We will study probe branes in this geometry to learn
more about the first order phase transition out of equilibrium.

The particular duality we concentrate on is the simplest example
of holography with fundamental quark fields
\cite{Polchinski,Bertolini:2001qa,Karch,Mateos,Babington,
Erdmenger:2007cm}. We do not consider the specific degrees of
freedom of the theory too crucial - it is some strongly coupled
gauge theory that displays a chiral phase transition. We hope, in
the spirit of AdS/QCD models \cite{Erlich:2005qh,DaRold:2005zs},
that it reflects broad aspects of many strongly coupled systems.
The specific gauge theory is constructed from the D3/D7 system in
type IIB string theory which we will describe further below. The
theory is the large $N_c$ ${\cal N}=4$ $U(N_c)$ gauge theory with
a small number of quark hypermultiplets. We will work in the
quenched approximation \cite{Karch} (appropriate when $N_f \ll
N_c$) which on the gravity dual side corresponds to treating the
D7 branes as probes in the metric generated by the D3 branes.
There is a $U(1)$ chiral symmetry (a remnant of the $SU(4)$
R-symmetry of the ${\cal N}=4$ theory) which is broken when a
quark condensate forms \cite{Babington, Kirsch2004,
Mateos:2006nu}. Several mechanisms for triggering this
condensation have been explored. The cleanest is when a background
magnetic field is introduced \cite{Johnson1,Erdmenger1,Zayakin,
Johnson2, Evans1} so we will use that mechanism here. The reader
might want to loosely view the B field as simply the introduction
of a conformal symmetry and supersymmetry breaking parameter that
triggers the strong dynamics to cause the symmetry breaking.
Physically, magnetic fields may be strong during structure
formation in the early universe, in particular during the epoch of
the QCD primordial phase transition, and in non-central
high-energy heavy ion collisions studied at RHIC. Running of the
coupling in the holographic dual also causes quark condensation as
has been shown in back-reacted dilaton flow geometries
\cite{Babington, Ghoroku:2004sp} and models with a
phenomenologically imposed dilaton profile \cite{Alvares:2009hv}.
The quark condensate can be determined in these models and an
effective IR quark mass is generated. The theories display a
massless pion-like Goldstone field and a massive sigma field
(since we are at large $N_c$ it is stable) that is the effective
Higgs particle.

The equilibrium finite temperature behaviour of the theory with a
magnetic field has been studied in \cite{Johnson1, Evans1}. Finite
temperature can be included through the presence of an AdS
Schwarzschild black hole. At a critical temperature the D7
embedding flips from a chiral symmetry breaking embedding away
from the horizon to a symmetry preserving embedding that enters
the horizon. The transition is therefore also associated with
meson melting \cite{Hoyos:2006gb} - for embeddings away from the
horizon there are regular linearized fluctuations describing the
meson spectrum of the theory. For the embedding that enters the
horizon there are only in-falling quasi-normal modes describing
unstable plasma fluctuations. In terms of the quarks of the theory
the high temperature phase is analogous to the quark gluon plasma
phase in QCD whilst the low energy phase is more akin to the
hadronic phase of QCD. It should be noted though that the gluonic
degrees of freedom deconfine at any finite temperature so the
analogy is imperfect.

The crucial extra ingredient we shall add to this story in this
paper is provided by the boost-invariant expanding or contracting
plasma geometry of Janik and Peschanski~\cite{Janik:2005zt}. This
geometry has a black hole whose horizon moves away from the boundary
in time as the
${\cal N}=4$ plasma it describes expands and cools. The time
reversed solution at zero viscosity describes a heating plasma and we will find it
useful to discuss that scenario too below. The geometry is a late
time expansion (when the black hole is small) in powers of inverse
time. However, by controlling the strength of the magnetic field
on the D7 probes, felt by the quarks, we can arrange to place the
chiral phase transition at any point in the evolution so the
expansion is sufficient to fully study the transition.

We will place a D7 brane\footnote{By placing a fundamental string,
which corresponds to a heavy quark, into the expanding plasma geometry,
the diffusion constant~\cite{Kim1} and drag force~\cite{Giecold} was computed.} 
into the expanding plasma geometry and
determine the partial differential equation (PDE) that describes
the time dependence of its embedding. A good first approximation
to the transition behaviour is provided by the equilibrium results
with the temperature replaced by the temperature as a function of
time from the moving background. In fact we will see that that is
an extremely good approximation when talking about the slow or
adiabatic heating or cooling of vacuum configurations at
temperatures even close to the phase transition. If the
chiral symmetry breaking embedding is heated though, the local
minima in the effective potential associated with that embedding
is eventually lost and the configuration becomes an out of
equilibrium configuration. The equilibrium results can not
describe the subsequent evolution. Similarly excited vacuum states
must be studied through the full PDE system.

We first turn to an approximation to the PDE. The solution can be
power expanded in inverse powers of time. This reduces the PDE to
a system of ordinary differential equations (ODEs) that are much
easier to solve. This analysis was first done in \cite{Grosse}
where the time evolution of the high temperature phase of the pure
${\cal N}=2$ theory was studied.  Using this technique we solve
the ODE system and find the moving D7 solution. This method
assumes that the initial vacuum state is exactly a maximum or
minimum of the zero temperature effective potential. Again the
heating is essentially adiabatic in nature. The expansion breaks
down if the super-heated state ceases to be an extrema of the
effective potential. This method allows us to confirm the success
of the equilibrium derived results although in fact the expansion
breaks down before the equilibrium results deviate from the full
PDE solutions.

The most interesting out of equilibrium questions lie beyond the
adiabatic approximation though. In a physical first order
transition quantities such as the condensate do not jump but the
vacuum state instead performs a fast roll from one configuration
to another. The timing of that transition can and most likely will
be spatially dependent \ie bubbles of the true vacuum will form
and grow. We turn to solving the full PDEs to study these
phenomena (with care this can be done using in built PDE solvers
in for example Mathematica). For example we are able to watch the
super-heated chiral symmetry breaking phase roll to the symmetric
phase. We can also simulate initial conditions with some extra
energy (which might for example come from thermal fluctuations)
and see such configurations transition away from the super-heated
vacuum before the meta-stable vacuum has disappeared in the
effective potential. This allows us to confirm some elements of
the transition such as the length of time in which there is a
mixed phase. We also are effectively watching a very large bubble
form.

Our main result then is to have developed numerical techniques
that let us reproduce the phase structure using the PDE solutions
and to describe non-equilibrium physics that is necessarily
present in the first order transition.

In our final section we also analyze the ODE expansion
approximation to the PDE solutions for the D7 embedding to make
clear the full dependence of the solutions on the magnetic field
value. In that case the dependence is available analytically. We
also show the effects of the viscosity.

\newpage

\section{Holographic Descriptions}

In this section we review the ${\cal N}=2$ gauge theory with a
magnetic field and its holographic description.  We discuss the
theory's finite temperature {\it chiral phase transition} in an
equilibrium description. We then review how to study flavour
physics in a {\it nonequilibrium} set-up using holography.

\subsection{Chiral symmetry transition} \label{21}

We will study the D3/D7 brane model at
finite temperature and with a magnetic field~\cite{Johnson1, Evans1}.
The magnetic field, which causes chiral symmetry breaking,
competes with the temperature that prefers to restore chiral symmetry.
Consequently there is a first order phase transition.

The ${\cal N} =4$ $SU(N)$ gauge theory at finite temperature has a
holographic description in terms of an AdS$_5$ black hole geometry which can be written
as
\begin{eqnarray}
   ds^2 = \frac{w^2}{R^2}(- g_t dt^2 + g_x d\vec{x}^2)
         + \frac{R^2}{w^2} (d\rho^2 + \rho^2 d\Omega_3^2
         + dL^2 + L^2 d\phi^2) \ , \label{metric2}
\end{eqnarray}
where $\phi$ is a $U(1)$ angle and
$w = \sqrt{\rho^2 + L^2}$, $\rho := w\sin\phi$, $L := w\cos\phi$ and
\begin{eqnarray}
g_t := \frac{(w^4 - w_H^4)^2}{2 w^4 (w^4+w_H^4)}\ ,  \qquad
g_x  := \frac{w^4 + w_H^4}{ 2 w^4} \ .
\end{eqnarray}
Note $R^4=4 \pi g_s N \alpha'{}^{2}$, and the temperature is given by $w_H := \pi R^2 T$.

Quenched ($N_f \ll N$) ${\cal N}$=2 quark superfields can be
included in the ${\cal N}=4$ $SU(N)$ gauge theory through probe D7 branes
in the geometry \cite{Karch}. The D7 probe can be
described by its Dirac-Born-Infeld (DBI) action
\begin{eqnarray}
S_{DBI} = - T_{D7} \int d^8\xi \sqrt{- {\rm det} (P[G]_{ab} +
2 \pi \alpha' F_{ab})} \ ,  \label{DBI}
\end{eqnarray}
where $P[G]_{ab}$ is the pullback of the metric and $F_{ab}$ is
the gauge field living on the D7 world volume. We will use
$F_{ab}$ to introduce a constant magnetic field~\cite{Johnson1},
\begin{eqnarray}
  F_{12} = -F_{21} = B/(2\pi\a') \ . \label{F12}
\end{eqnarray}
We embed the D7 brane assuming only $\rho$ dependence: $L(\rho)$
at constant $\phi$. The full DBI action we will consider is then
\begin{eqnarray}
  S_{DBI} = \int d\xi^8 \call(\rho)
    = \left(\int_{S^3} \e_3 \int dtd\vec{x} \right) \int d\rho \
  \call(\rho) \ ,
\end{eqnarray}
where $\e_3$ is a volume element on the 3-sphere and
\begin{eqnarray}
  -\call \equiv \wt{\Omega} \equiv N_f T_{D7} (R\sqrt{B})^4  \trho^3\left(1-\frac{\tw_H^4}{\tw^4}\right)
   \sqrt{\left(1+\tL'^2\right)} \sqrt{\left(\left(1+\frac{\tw_H^4}{\tw^4}\right)^2 + \frac{1}{\tw^4} \right)} \  \label{OriginalAction}
\end{eqnarray}
with the dimensionless variables defined as
\begin{eqnarray}
    (\tw , \tL, \trho)  := \left(\frac{w}{R\sqrt{2B}},
        \frac{L}{R\sqrt{2B}} ,
       \frac{\rho}{R\sqrt{2B}} \right) \ .
\end{eqnarray}
Note that we can use the magnetic field value as the intrinsic
scale of conformal symmetry breaking in the theory - that is we
can rescale $L$ and $\rho$ by $B$. The Euclidean on-shell
Lagrangian ($-\call$) is interpreted as the free energy density
($\wt{\Omega}$).

\begin{figure}[]
\centering
  \subfigure[\label{}]
  {\includegraphics[width=5.cm]{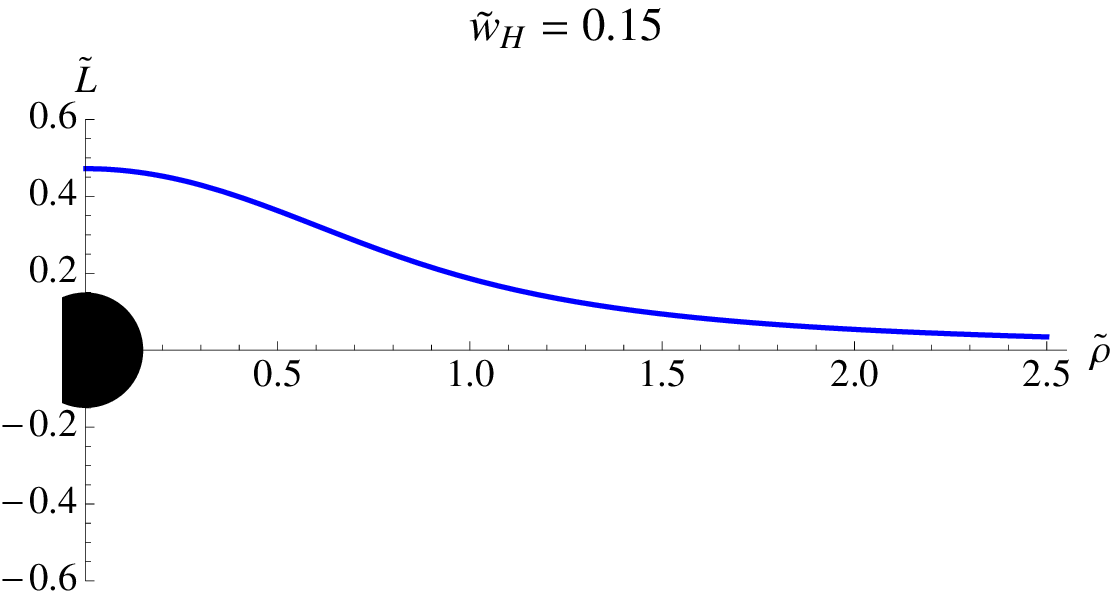}
   \includegraphics[width=5.cm]{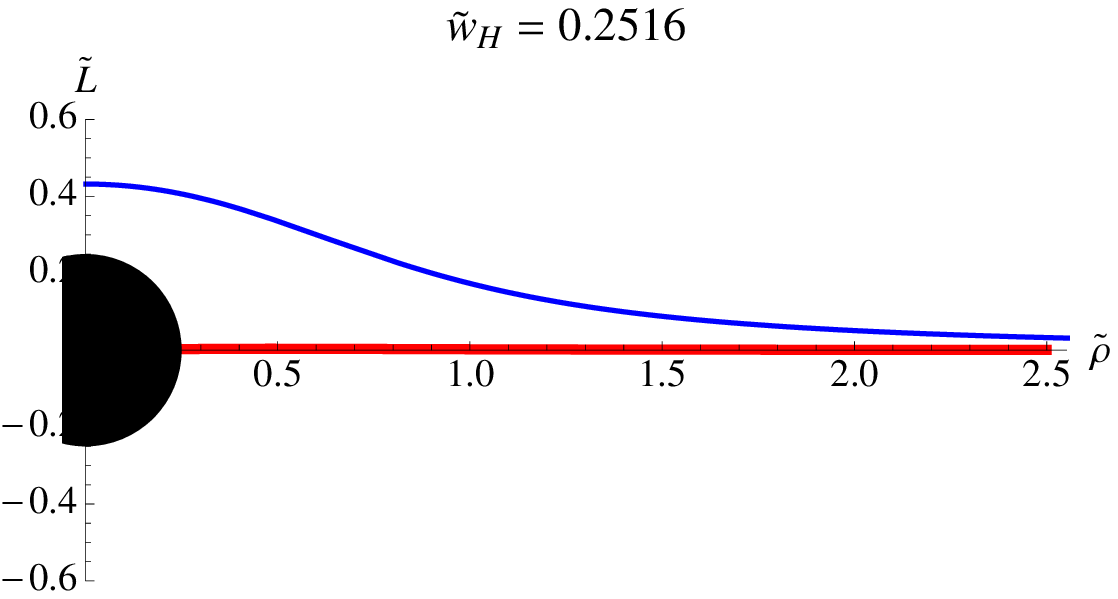}
   \includegraphics[width=5.cm]{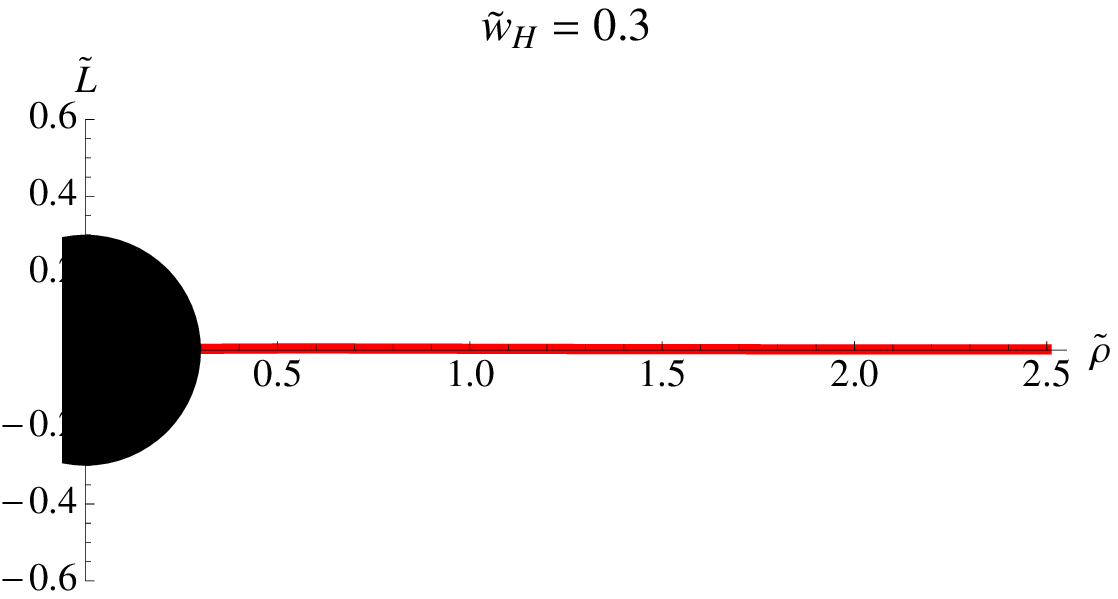}}
   \subfigure[\label{}]
   {\includegraphics[width=6cm]{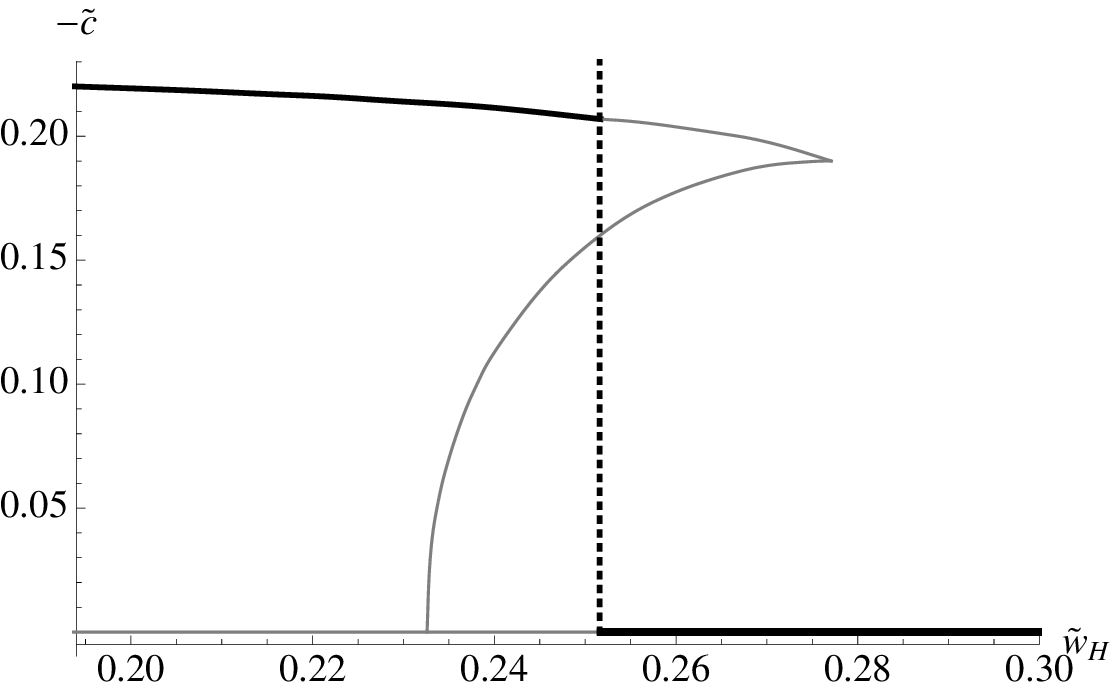}}
  \caption{
            The equilibrium description of the first order chiral phase transition in the ${\cal N}=2$ gauge theory
            with a magnetic field: (a) The D7 brane embedding profiles $\tilde L(\trho)$
            as a function of temperature $\tw_H$. The low
            temperature symmetry breaking embedding is shown on the left
            whilst the symmetric flat embedding on the right is
            preferred at high temperature.
            (b) A plot of the chiral condensate as a function of the temperature with the dotted
            line indicating the position of the first order transition.
           }\label{Fig1}
\end{figure}

In all cases the embeddings become flat at large $\rho$  taking
the form
\begin{eqnarray}
  \tL(\trho) \sim \tm + \frac{\tc}{\trho^2}\ , \quad
  \tm = \frac{2\pi \a' m_q}{R\sqrt{2B}} \ , \quad
  \tc = \brac{\bar{q}q} \frac{(2\pi \a')^3}{(R\sqrt{2B})^3} \ , \ \label{scaled}
\end{eqnarray}
where $\tm$ and $\tc$ are identified with the quark mass
and the quark condensate, respectively. 
Since we are interested in spontaneous symmetry breaking we impose
$\tm=0$ in the UV ($\trho \ra \infty$). Then the value of $\tc$ is
determined by requiring regularity in the IR ($\trho \ll 1$). If
there is more than one solution then we choose the one giving the
minimum free energy.

The results are displayed in Fig.~\ref{Fig1}.
At low temperatures $T\ll B$, the black hole is small and the embeddings are
repelled from the origin of the $\tL-\trho$ plane (Fig.~\ref{Fig1} (left)).
This behaviour is a result of the inverse
powers of $\tw^4$, when $\tw_H \ll 1 $, in the last term of the Lagrangian
(\ref{OriginalAction}), which causes the action to grow if the D7 approaches
the origin. Consequently $\tc$ is non-zero and chiral symmetry is broken.
If the temperature is allowed to rise sufficiently then the black
hole horizon grows to mask the area of the plane in which the
inverse $\tw^4$ terms in the Lagrangian are large. At a critical
value of $T$ ($\tilde{w}_H = 0.2516$, Fig.~\ref{Fig1} (middle))
the benefit to the $\tm=0$ embedding of curving off
the axis becomes disfavoured and it instead lies along the
$\tilde{\rho}$ axis - chiral symmetry breaking switches off.
This transition is first order since the condensate vanishes
discontinuously, which corresponds to the embedding change from the blue to the
red one.
At higher temperatures (Fig.~\ref{Fig1} (right)),
the embedding stays flat, $\tc=0$, as expected in the chirally symmetric phase.
The corresponding condensate vs.\ temperature ($-\tilde{c}$-$\tilde{w}_H$) plot is shown in 
Fig.~\ref{Fig1} (bottom).

\subsection{A boost-invariant expanding plasma} \label{22}

The geometry (\ref{metric2}) is dual to a system in thermodynamical
equilibrium and therefore not suitable for the description of
the chiral phase transition in a rapidly expanding plasma, in which
the transition is basically a non-equilibrium process.
Boost-invariant expanding ${\cal N} =4$ $SU(N)$ plasmas
out of equilibrium can however be described by the time-dependent
background found in \cite{Janik:2005zt} 
 (see \cite{Janik:2010we} for a review). In the following we review
the basic features of this background and discuss the embedding of probe D7 branes
dual to quenched flavours (``quarks'') in the plasma.

The boost-invariant geometry is a 5D spacetime with coordinates
$\{\t, y, x_\perp (= x^1, x^2), z\}$, which, apart from the holographic
direction $z$, parameterise the 4D
spacetime on the boundary. The longitudinal position plane
is parameterised by the proper time $\tau$ and rapidity $y$
(related to $x^{0,3}$ as $x^0 = \t \cosh y, \ x^3 = \t \sinh y$),
the transverse coordinates are collected in $x_\perp$. We also add
a five-sphere to obtain a full type IIB supergravity background.
The metric is then of the form
\begin{eqnarray}
  \frac{ds^2}{R^2} = \frac{1}{z^2}\left( - e^{a(\t,z)} d\t^2 + e^{b(\t,z)} \t^2 dy^2
  + e^{c(\t,z)} dx_\perp^2   \right) + \frac{dz^2}{z^2} + d\Omega_5^2 \ , \label{BGmetric}
\end{eqnarray}
where $R$ is the radius of the AdS$_5$ space. At late times, the coefficients can be
expanded to first order  as~\cite{Janik:2005zt, Nakamura, Janik:2006ft}
\begin{eqnarray}
  && a(\t,z) = \ln \left(\frac{(1-v^4/3)^2}{1+v^4/3}\right)
          +   2 \eta_0 \frac{(9+v^4)v^4}{9-v^8} \left[ \frac{1}{(\ve_0^{3/8}\t)^{2/3}} \right]
          + \calo\left[\frac{1}{\t^{4/3}}\right]  , \nn \\
  && b(\t,z) = \ln (1 + v^4/3 )
          + \left( - 2 \eta_0 \frac{v^4}{3+v^4} + 2\eta_0 \ln \frac{3-v^4}{3+v^4}\right)
          \left[\frac{1}{(\ve_0^{3/8}\t)^{2/3}}\right]
          + \calo\left[\frac{1}{\t^{4/3}}\right]  , \, \qquad  \label{abc}  \\
  && b(\t,z) = \ln (1 + v^4/3 )
          + \left( - 2 \eta_0 \frac{v^4}{3+v^4} - \eta_0 \ln \frac{3-v^4}{3+v^4}\right)
          \left[\frac{1}{(\ve_0^{3/8}\t)^{2/3}}\right]
          + \calo\left[\frac{1}{\t^{4/3}}\right]   ,\nn
\end{eqnarray}
with
\begin{eqnarray}
  v \equiv \frac{z}{\t^{1/3}} \ve_0^{1/4} \ , \qquad \eta_0 = \frac{1}{2^{1/2}3^{3/4}} \ .
  \label{veta}
\end{eqnarray}
$\ve_0$ is a free parameter of mass dimension $8/3$ and is related
to the energy density, while $\eta_0$ is related to the shear
viscosity. $v$ is a scaling parameter valid at large $\t$. Note
that $a(\t,z),b(\t,z)$ and $c(\t,z)$ are expanded around $\t =
\infty$ in powers of $1/\t^{2/3}$
as~\cite{Janik:2005zt, Nakamura, Janik:2006ft}
\begin{eqnarray}
  a(\t,z) = \sum_{n=0}^\infty a_n(v) \left( \frac{1}{\ve_0^{3/8}\t} \right)^{\frac{2}{3}n} \ , \label{abc0}
\end{eqnarray}
and similarly for $b(\t,z)$ and $c(\t,z)$. The coefficients are
functions of a scaling parameter $v$ only. Due to this scaling
behaviour the complicated PDE Einstein equations can be reduced to
an ODE system, which allows analytic solutions. At early time $\t
\ll 1$ there is no scaling behaviour and we should solve the full
PDEs.

To find which gauge theory state corresponds to this bulk metric
according to the AdS/CFT dictionary let us expand the metric
around $z=0$,
\begin{eqnarray}
  && g_{\t\t} \equiv  -e^{a(\t,z)} = -1 + \frac{2\pi^2}{N_c^2} z^4 \left[ \bar{\ve}
              - \frac{2\eta}{\t} \right] + \calo(z^6) \,, \nn \\
  && g_{yy} \equiv  \tau^2 e^{b(\t,z)} = \t^2  + \frac{2\pi^2}{N_c^2} z^4
    \left[ \t^2 \left( \frac{\bar{\ve}}{3}
   -\frac{2\eta}{\t}\right)\right] + \calo(z^6) \,, \label{Bmetric} \\
  && g_{11} (= g_{22}) \equiv e^{c(\t,z)} =
  1 + \frac{2\pi^2}{N_c^2} z^4 \left[\frac{\bar{\ve}}{3}\right] + \calo(z^6) \nn  \ ,
\end{eqnarray}
where \footnote{$\frac{2\pi^2}{N_c^2}$ has been factored out to
apply the AdS/CFT dictionary
\begin{eqnarray}
  \langle T_{\m\n} \rangle = \frac{N_c^2}{2\pi^2} \lim_{z\ra0} \frac{1}{z^4}(g_{\m\n}-\eta_{\m\n})
   \,.
\end{eqnarray}
}
\begin{eqnarray}
  \bar{\ve} \equiv \frac{N_c^2}{2\pi^2}\frac{\ve_0}{\t^{4/3}} \ ,  \qquad
  \eta \equiv \frac{N_c^2}{2\pi^2}\frac{\eta_0 \ve_0^{3/4}}{\t} \ .
\end{eqnarray}
The leading terms (of order $\calo(z^0)$) of the metric elements (\ref{Bmetric})
is simply the Minkowski metric in the $\t - y$ coordinate system, and the
subleading terms (of order $\calo(z^4)$) correspond to the expectation value
of the energy-momentum tensor, \ie
\begin{eqnarray}
 \langle T_{00} \rangle =  \bar{\ve}-\frac{2\eta}{\t} \ , \quad
 \langle T_{yy} \rangle =  \t^2 \left(\frac{\bar{\ve}}{3}-\frac{2\eta}{\t}\right) \ , \quad
 \langle T_{11} \rangle = \langle T_{22} \rangle
  =  \frac{\bar{\ve}}{3} \ . \quad
\end{eqnarray}
This energy-momentum tensor is precisely that of a longitudinal
viscous boost-invariant $\caln=4$ SYM conformal plasma
with finite $\eta$.

In $g_{\t\t}$ we recover a time-dependent emblackening factor, which describes
a moving horizon. The size of the horizon determines the time-dependence
of the `temperature' as~\cite{Nakamura,Grosse}
\begin{eqnarray}
T(\t) = \frac{\sqrt{2}}{\pi R^2} r_H = \left( \frac{4\ve_0}{3}
\right)^{1/4}\frac{1}{\pi\t^{1/3}} \left(
1-\frac{\eta_0}{2\ve_0^{1/4}\t^{2/3}}   \right) \ . \label{Temp}
\end{eqnarray}
If we assume that the time-dependent entropy density $s(\t)$ has the same form
as in the static case, $s(\t) = (\pi^2/2)N_c^2 T(\t)^3$, then the ratio
$\eta/s$ can be computed as~\cite{Janik:2006ft}
\begin{eqnarray}
  \frac{\eta}{s} = \frac{1}{4\pi} + \calo(\t^{2/3}) \ ,
\end{eqnarray}
which at large $\tau$ agrees with the known static bound. Note that the numerical
value of $\eta_0$ in (\ref{veta}) is crucial to get $1/4\pi$.

Finally we note the uniqueness of the gravity solution. There is a
potential singularity at $v=3^{1/4}$ in (\ref{abc}). The curvature
invariant, $ R_{\m\n \rho \s} R^{\m\n \rho \s} = R_0(v) +
R_1(v)\t^{-2/3} + R_2(v)\t^{-3/4} + \cdots $ is only regular at
each order if one makes the specific choices above. We must choose
$-1/3$ for the power of $\t$ in (\ref{veta}) to make $R_0$
regular~\cite{Janik:2005zt}. We must also choose (\ref{abc})
with the specific choice of the numerical value
$\eta_0$ as in (\ref{veta}) to make $R_2$
regular~\cite{Janik:2006ft}. ($R_1$ is always regular and does not give any constraint.)

\medskip
Flavours in this background were first studied in \cite{Grosse}.
Knowing the explicit background geometry (\ref{BGmetric}) we can
study flavour physics using the D7 brane DBI action. The action
schematically reads
\begin{eqnarray}
  S  \sim  \int d^8\xi \sqrt{- {\rm det} (P[G]_{ab}}) \sim
  \int d\t d\rho\, \t \rho^3 \mathbb{A}
  \sqrt{
  1+(\del_\rho L)^2-\mathbb{B} \frac{ (\del_\t L)^2}{(\rho^2 + L^2)^2}  }
  \,,\label{simpleaction}
\end{eqnarray}
where $\mathbb{A}$ and $\mathbb{B}$ are complicated but known functions of $\t$ and $\rho$,
see (\ref{A}) and (\ref{B}) in the next section. 
$L=L(\tau,\rho)$ is the embedding profile of the D7 brane and is assumed to be
a function of $\t$ and $\rho$.

The equation of motion coming from (\ref{simpleaction}) is a
non-linear PDE. However it can be semi-analytically solved by the
late-time expansion
\begin{eqnarray}
  L(\tau, \rho) = m + \sum_{i=1}^{\infty}{f}_i(\rho) \tau^{-\frac{i}{3}} \label{Ansatz1} \,.
\end{eqnarray}
This expansion inherently assumes that the late time configuration is precisely the equilibrium vacuum
and when we heat it there is no excess energy. This leads to an adiabatic approximation.
The fraction $1/3$ is chosen because all exponents of $\t$ in
$\mathbb{A}$ and $\mathbb{B}$ are integer multiples of $1/3$
and the constant $m$ reflects the fact that the embeddings get flat at large $\tau$.
Here we only consider a large bare quark mass ($m \gg 1$) and Minkowski embeddings,
{\em i.e.}\ D7 configurations which end well above the black hole horizon.

This reduces the partial differential equation to an infinite set of ordinary
differential equations given by
\begin{eqnarray}
  && \rho^{-3} \del_\rho \left( \rho^3 {f}'_i(\rho) \right) = \frac{8m\ve_0^2 }{9(m^2+\rho^2)^5} I_i \ ,   \label{EOM1} \\
  && I_8 = 1 \ , \quad  I_{10} = -4\eta_0\ve_0^{-1/4}\ , \quad  \mathrm{otherwise}\ \  I_{i \le 13} = 0  \ ,
\end{eqnarray}
where we do not consider terms with $i > 13$ since they are beyond the validity
regime of our approximation of the boost-invariant metric.
The asymptotic solution at large $\rho$ is
\begin{eqnarray}
  {f}_i \sim {m}_i + \frac{{c}_i}{\rho^2} \label{Asymp1}\,.
\end{eqnarray}
We set ${m}_i = 0$ so the bare quark mass is zero and not time-dependent.
We also impose the condition $f'_i(0) = 0$ for regularity.
With these boundary conditions, we find $f_i = 0$ except for $f_8$, $f_{10}$.
As a result, one gets
\begin{eqnarray}
  L(\tau, \rho) = m + c(\tau) \frac{\rho^4 + 3\rho^2 m^2 + 3 m^4}{(m^2 + \rho^2)^3}  \label{L1} \,,
\end{eqnarray}
where
\begin{eqnarray}
  c(\tau) = - \frac{\e^2_0}{54 m^5} \left( \t^{-8/3} -4\eta_0 \e_0^{-1/4}\t^{-10/3} + \cdots  \right) \ .
  \label{c1}
\end{eqnarray}
The condensate approaches zero as $\sim - \ \t^{-8/3}$ and the viscosity has a ``dragging" effect
$\sim + \  \t^{-10/3}$.

We close this section with some technical remarks on black hole
embeddings.  In the Fefferman-Graham (FG) coordinate system
(\ref{BGmetric}) one cannot approach the horizon for fixed $\tau$
as in the Schwarzschild black hole metric.  There is also an
extended background written in terms of Eddington-Finkelstein (EF)
coordinates, where the spacetime is well defined across the
horizon~\cite{Heller:2008mb,Kinoshita:2008dq} and black hole
embeddings may be described more consistently there.  However we
have found that the embedding configuration is not easy to handle
in those coordinates.  Below, to enable us to study black hole
embeddings, we use FG coordinates but with a cut-off slightly
above the horizon. Since the FG coordinate system is a good patch
near the horizon at large $\tau$, by restricting ourselves to
large $\tau$, we may capture the essential physics of the
embedding. This will allow us to go beyond the results in
\cite{Grosse}.

\section{Out of equilibrium description of the chiral phase transition}

In the previous section we reviewed the holographic description
of dynamical flavours in an expanding plasma \cite{Grosse}.
A~boost-invariant background with embedded D7 branes is however not sufficient
to study the chiral transition. In that case the D7 embeddings are
always flat in the chiral limit (\mbox{$m\rightarrow 0$}) and the system is
always in the chiral symmetric phase with vanishing quark condensate.
 In order to describe the transition to the chiral
broken phase, we need a {\it repulsive} effect to compete against
the attractive force of the black hole. As in the static case
reviewed in section \ref{21}, this can be achieved by turning on a
magnetic field. In this section, we will therefore consider D7
branes with a world-volume magnetic field in the dual geometry of
an expanding plasma. In this way we will find the holographic dual
of a chiral transition and deduce the dynamic effective potential
for the time-dependent quark condensate.

\subsection{D7 flavour brane action}

The background metric for a boost-invariant expanding plasma
can be written as
\begin{eqnarray}
ds^2 = \frac{r^2}{R^2} (- e^{a(\t,r)} d\tau^2 + e^{b(\t,r)} \tau^2 dy^2
+ e^{c(\t,r)}dx_{\perp}^2)
+ \frac{R^2}{r^2} (d\rho^2 + \rho^2 d\Omega^2_3 + dL^2 + L^2 d\phi^2) \quad
\end{eqnarray}
where $r^2 \equiv \rho^2+L^2$. The $S^5$ part is written as in
(\ref{metric2}) and the AdS$_5$ part follows from (\ref{BGmetric})
with a change $z \ra R^2/r$, {\em i.e.}\  $a(\t,r) \equiv
a(\t,z\ra R^2/r)$  in (\ref{abc}) and similarly for $b(\t,r)$ and
$c(\t,r)$.

We are interested in time-dependent D7 brane embeddings
of the type $L=L(\tau,\rho)$. The corresponding DBI action
is
\begin{eqnarray}
S_{DBI} = - T_{D7} \int d^8\xi \sqrt{- {\rm det} (P[G]_{ab} +
2 \pi \alpha' F_{ab})} \ ,
\end{eqnarray}
where we turn on a constant magnetic field~\cite{Johnson1},
\begin{eqnarray}
  F_{12} = -F_{21} = B/(2\pi\a')  \ , \label{F122}
\end{eqnarray}
in order to induce the chiral symmetry breaking.

More explicitly, the D7-brane action reads
\begin{eqnarray} \label{pdeaction}
  &&S = \mathbb{N} \int d\t d\rho\, \t \rho^3 \mathbb{A}
  \sqrt{\left(1+\mathbb{C}\frac{R^4 \nB^2}{(\rho^2+L^2)^2}\right)
  \left(1+L'^2-\mathbb{B} \frac{R^4 \dot{L}^2}{(\rho^2 + L^2)^2}  \right)} \label{action}\\
  &&\qquad \mathbb{A} \equiv \left(1-\frac{v^8}{9}\right)
  \mathrm{exp}\left[4 \eta_0 \ve_0^{-1/4} \frac{v^8}{9-v^8} \t^{-2/3} \right]    \ , \label{A} \\
  &&\qquad \mathbb{B} \equiv \frac{1+\frac{v^4}{3}}{\left(1-\frac{v^4}{3}\right)^2}
  \mathrm{exp}\left[-2 \eta_0 \ve_0^{-1/4} v^4 \frac{9+v^4}{9-v^8} \t^{-2/3} \right]  \label{B}  \ ,  \\
  &&\qquad \mathbb{C} \equiv \frac{9}{\left(3+v^4\right)^2}
  \mathrm{exp}\left[4 \eta_0 \ve_0^{-1/4} \left(\frac{v^4}{3+v^4}
  - \mathrm{Coth^{-1}}\left(\frac{3}{v^4}\right)\right) \t^{-2/3} \right]  \ , \label{C} \\
  &&\qquad v \equiv \frac{\ve_0^{1/4} R^2}{\t^{1/3} \sqrt{\rho^2 + L^2}}  \ ,
   \quad \mathbb{N} \equiv N_f T_{D7} \int \epsilon_3 \int dy d^2 x_\perp \ ,
\end{eqnarray}
where ${\mathbb A}=e^{a/2+b/2+c}$, $
{\mathbb B}=e^{-a}$, ${\mathbb C} = e^{-2c}$, $\eta_0 = 1/{ (2^{1/2} 3^{3/4})} $
as in (\ref{veta}),
and $\epsilon_{3}$ is the volume form of the three-sphere.
For vanishing $\nB$, the action reduces to that in (\ref{simpleaction}).

It turns out to be convenient to work with the rescaled
variables
\begin{eqnarray}
  \rho \equiv \sqrt{\nB} R \trho \ , \quad L \equiv \sqrt{\nB}R \tL \ ,
  \quad \t \equiv \frac{R}{\sqrt{\nB}}\ttau
   \ , \quad \ve_0 \equiv \nB^{4/3} R^{-8/3} \wt{\ve}_0 \label{Dimless}\, \ .
\end{eqnarray}
For $R=1$ the action then reads
\begin{eqnarray}\label{action2}
  S = \mathbb{N} \nB \int d\ttau d\trho\, \ttau \trho^3 \mathbb{A}
  \sqrt{\left(1+\mathbb{C}\frac{1}{(\trho^2+\tL^2)^2}\right)
  \left(1+\tL'^2 -\mathbb{B} \frac{ \dot{\tL}^2}{(\trho^2 + \tL^2)^2}  \right)} \ ,
\end{eqnarray}
with $\mathbb{A,B,C}$ as in (\ref{A})-(\ref{C}) but
now expressed in terms of $\ttau, \wt{\ve}_0, \trho$, and $\tL$.

\subsection{Naive equilibrium based approximation}

\begin{figure}[]
    \centering
    \subfigure[]
   {\includegraphics[width=7.cm]{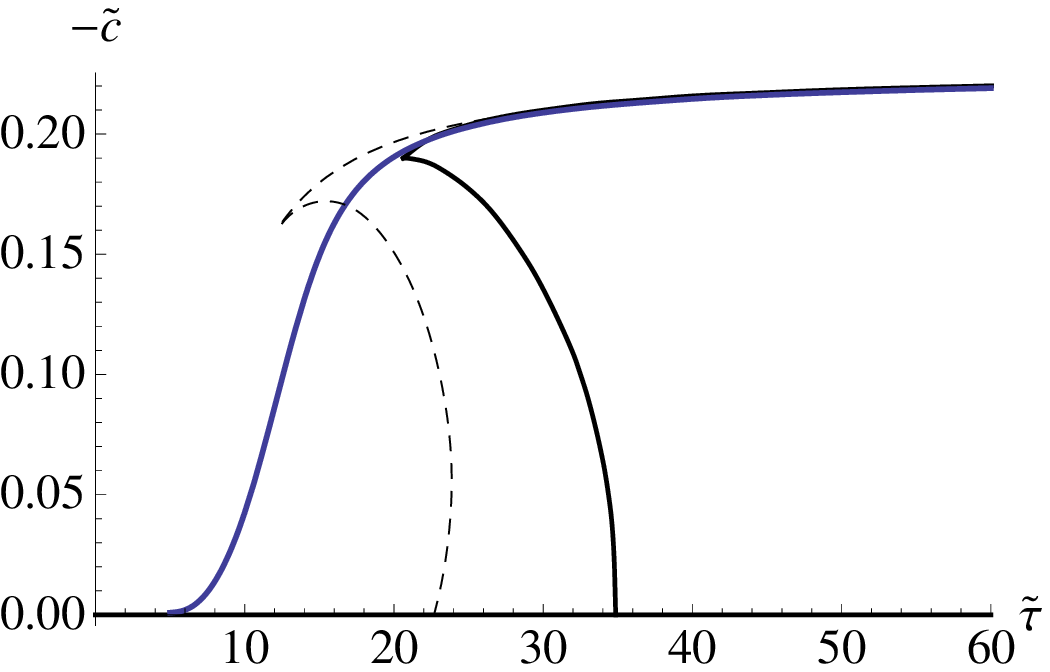}}{\hspace{0.5cm}} \hspace{-0.5cm}
    \subfigure[ ]
   {\includegraphics[width=7.5cm]{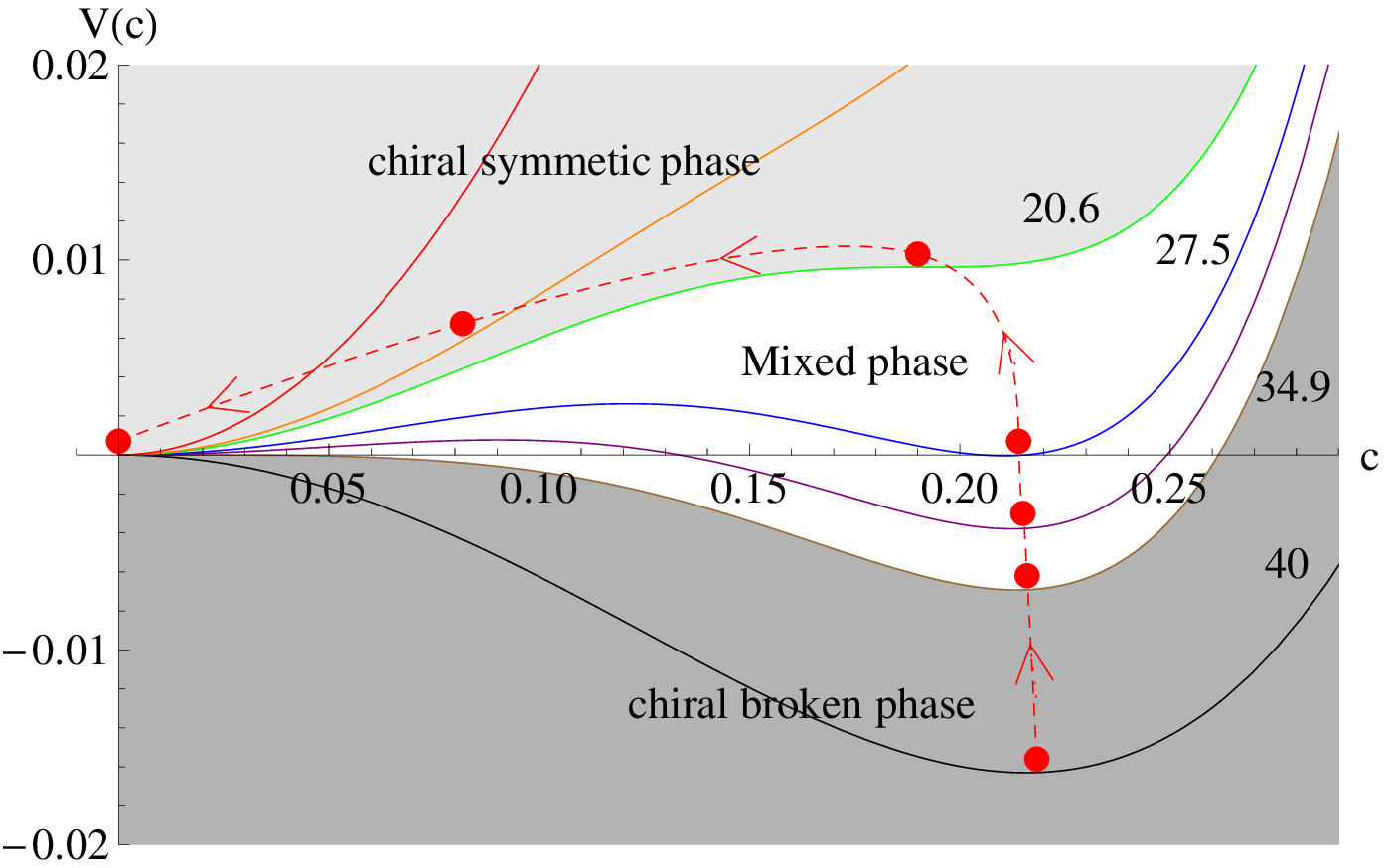}}
        \caption{\label{Fig.c} A summary of our main results showing the first order phase transition in our
        various approaches. (a) The condensate $\tc(\ttau)$ in the equilibrium (black), adiabatic
    (dashed) and non-equilibrium (blue) approaches.
    In the equilibrium description, the different branches correspond to the extrema of the potential $V(c)$ (b).  The complete potential is obtained by a $U(1)$ rotation around the vertical axis. 
    The dashed red curve schematically shows the path of the adiabatic heating evolution.}
\end{figure}

An obvious first approximation to understanding the time dependent
chiral phase transition in this set up is to use the equilibrium results from
section 2.1. There we described how the D7 embedding behaved in the
background of a fixed size black hole and determined the quark
condensate as a function of temperature. In the cooling plasma geometry
of section~2.2 the black hole horizon moves as a function of time $\ttau$ as
\begin{align}
{\tilde r}_H = \frac{\te_0^{1/4}}{3^{1/4} \ttau^{1/3}}
\left( 1-\frac{\eta_0}{2\tilde \ve_0^{1/4}\tilde \t^{2/3}}   \right)\,.  \label{rtau}
\end{align}
If this heating were very slow (as it is at large $\ttau$) we would expect to be able to
plot the quark condensate against $\ttau$ by simply substituting for the
temperature $T$ in the equilibrium results. We show that plot in Fig.~\ref{Fig.c}a
(black solid curve) which follows directly from the $c$-$T$ plot in Fig.~\ref{Fig1}.

We can recast the $\tc$ vs.~$\ttau$ plot schematically as a time
dependent effective potential by fitting it at each value of
$\ttau$ to a potential of the form
\begin{eqnarray} V = m^2 \tc^2 + \lambda \tc^4 + \l' \tc^6 \end{eqnarray}
using the values of $\tc$ at the extrema to fix the parameters
(the overall scale is not set but the figure is intended to only
be schematic). We plot this in Fig.~\ref{Fig.c}b. This is the
standard picture of a first order transition.

It is important to interpret these plots correctly. Firstly note
that there is a $\tc=0$ solution for all $\ttau$. If we begin at
high temperature with the symmetric flat embedding and cool slowly
or adiabatically (\ie move to large $\ttau$) then we can remain in
that embedding for all $\ttau$.  Above the time $\ttau_a=20.6$ two
extra solutions for the condensate develop and these, as we will
discuss, trigger the first order transition. At large $\ttau$ the
$\tc=0$ flat embedding becomes a local maximum of the effective
potential. To stay in the flat embedding for all~$\ttau$ is the
extreme limit of super-cooling the high temperature phase into an
unstable vacuum state.

The top trajectory in the $\tc$ vs.~$\ttau$ plot is best thought
about in the time reversed solution that is heating up. At early
times (large $\ttau$ in the plot) the solution is the $T=0$
symmetry breaking D7 embedding vacuum. Now as we heat
adiabatically the condensate tracks along the trajectory to
smaller $\ttau$. The solution ceases to exist at $\ttau_a$, where
the cusp in the (black) curve is, indicating that the minimum of
the effective potential corresponding to this solution has ceased
to exist at $\ttau_a$. At this point the brane will move quickly,
as an out of equilibrium configuration, ending as the flat
embedding or oscillating about it. The real-time evolution of the
configuration is represented by the path of red dots in Fig.~2b.
Starting from the global minimum with $c\neq 0$, the minimum is
lifted up and the configuration moves adiabatically. At $\ttau_a$
the minimum disappears and the system roles down the potential to
the true vacuum with $c=0$. The equilibrium results can not tell
us about this motion.

Now we also see the role of the middle arc in the plot - these
solutions are symmetry breaking embeddings that end on the horizon
and they are a local maximum of the potential lying between the
two minima (the embeddings of the top trajectory and the flat
embedding).

In fact the first order transition point between the symmetry
preserving and the symmetry breaking embeddings can be computed
from the free energy in the equilibrium computation. This
transition occurs at $\ttau_c=27.5$, where the local
and the global minimum in the effective potential interchange their
roles, \ie for $\ttau_c \geq 27.5$ the symmetry breaking vacuum (with $c\neq 0$)
becomes the global minimum. Another important time
determined by the equilibrium computation is $\ttau_b=34.9$ where
the mixed phase ends at large $\ttau$.

Our first  task in solving the PDEs for the D7 brane motion that
result from (\ref{pdeaction}) is to demonstrate that this
description is essentially correct - we will see that it is. We
will try to find confirmation of the times $\ttau_{a-c}$ of these
events. The more interesting task is that we will be able to
follow the evolution of a particular initial condition through the
phase transition. Of course in the first order transition the
brane configuration does not discontinuously leap between the
symmetry breaking embedding to the flat embedding but evolves
continuously. We will provide solutions for this evolution.

Another interesting phenomena  associated with a first order
transition is bubble formation. In real systems thermal energy
will lead to volumes of space which have more energy than an
equilibrium like state in a local minimum of the potential at any
particular time. These volumes may ``climb" over the potential
hill to the other local minimum during the mixed phase period
shown in Fig.~\ref{Fig.c}. These bubbles then grow or contract
triggering the phase transition around $\ttau_c$ ending any
super-heated or cooled phase. We will not look at ($x_3$) spatial
dependent brane embeddings.  However, we can inject kinetic energy
in the holographic directions of our description into the brane
configuration before we heat or cool it to see the configurations
moving more quickly between the two local minima than the lowest
energy configuration. This will allow us to test the time period
in which the mixed phase exists in the out of equilibrium problem.

\subsection{Adiabatic dynamic D7 brane embeddings} \label{secadiabatic}

Our first study of the PDEs describing the D7 embedding in the expanding plasma
geometry will be to study adiabatic expansion.
The non-linear partial differential equations resulting from
(\ref{action2}) can again be transformed into a set of
second-order ordinary differential equations as we reviewed for the case with
no symmetry breaking in section 2.2. For that, we use the
late-time expansion
\begin{eqnarray}
 \tL(\ttau, \trho) = f_0(\trho) + \sum_{i=1}^{\infty}f_i(\trho)
 \ttau^{-\frac{i}{3}} \,. \label{factorout}
\end{eqnarray}
Thus we will be following the evolution of the symmetry breaking
embedding as $\tau$ decreases. There is also the solution $f_i=0$
which corresponds to the symmetric embedding. Note that, contrary
to (\ref{Ansatz1}), a nontrivial asymptotic embedding $f_0(\trho)$
is assumed because of the repelling effect of the magnetic field.
The equations for every $f_i(\trho)$ can be obtained order by
order in $\tau^{-\frac{1}{3}}$, which will be solved recursively.
Unlike (\ref{EOM1}) these equations are quite lengthy and will not
be presented here. They do not allow for an analytic solution such
as (\ref{L1}) but can be numerically solved without any
difficulty. As in the static case \cite{Babington}, there are
expected to be two types of brane solutions, Minkowski and black
hole embeddings, depending on whether the brane ends on or above
the horizon.

\begin{figure}[]
    \centering
    \subfigure[\label{Fig.fa}]
    {\includegraphics[width=6.5cm]{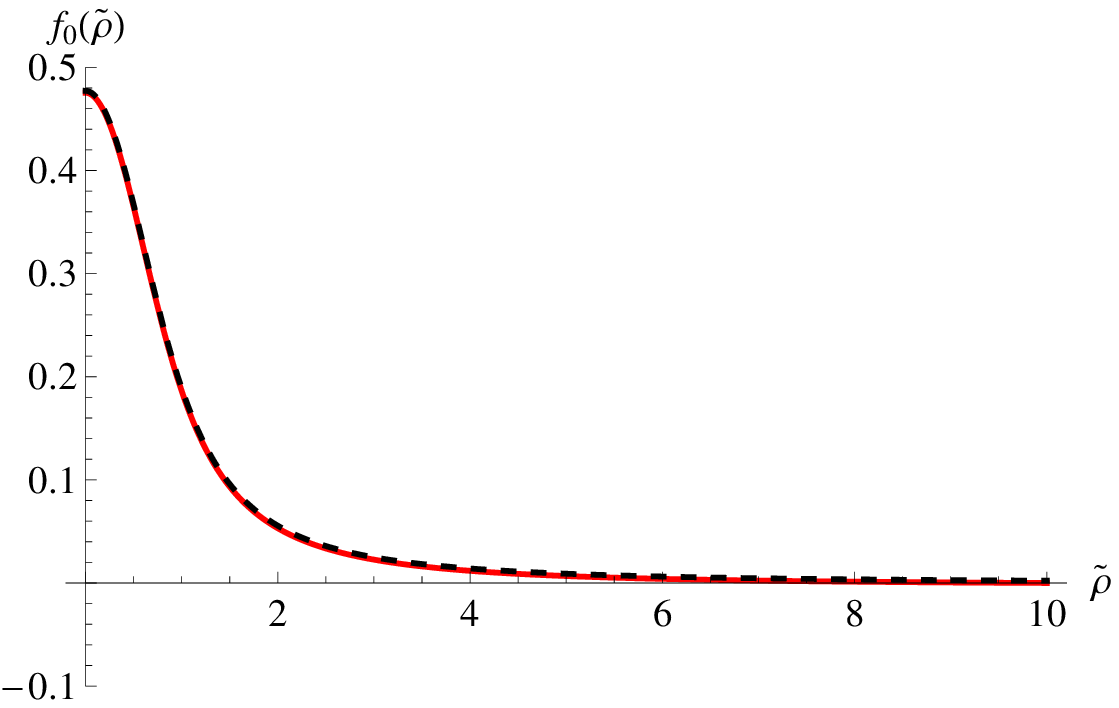}} \hspace{1cm}
     \subfigure[\label{Fig.fb}]
    {\includegraphics[width=6.5cm]{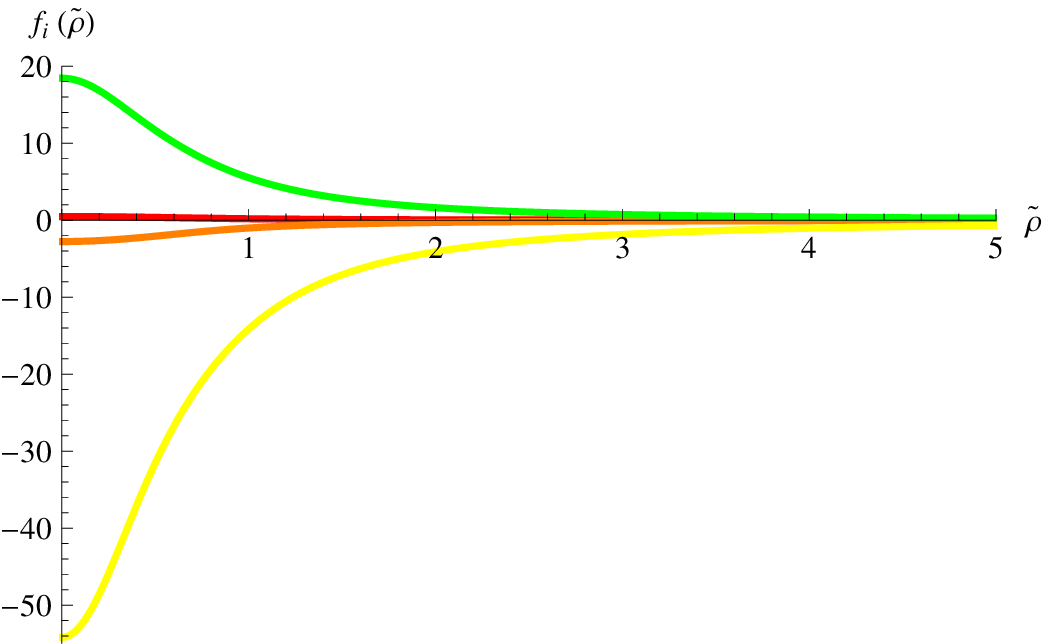}}
\caption{\label{Fig.f} Solutions of the ODEs in the adiabatic
approximation - $f_{0,4,8,10}(\trho)$ for $\te_0 = 1$. (a) $f_0$
 (red) compared to the dashed curve which is the profile
obtained in the static case at $T=0$. (b) $f_0$ (red), $f_4$
(orange), $f_8$ (yellow), $f_{10}$ (green).}
\end{figure}

For Minkowski embeddings, the boundary conditions are
\begin{align}
f_i'(0) = 0\qquad \textmd{and}\qquad f_i(\infty) = 0 \,,  \label{asstatic1}
\end{align}
as before. With these boundary conditions, we find non-trivial
profiles $f_i=f_i(\trho)$ for $i=0,4,8,10$, and $f_i=0$ otherwise. A
general dynamical embedding function is therefore of the form
\begin{eqnarray}
  \tL(\ttau, \trho) = f_0(\trho) + \frac{f_4(\trho;\te_0)}{\ttau^{4/3}}
  + \frac{f_8(\trho;\te_0)}{\ttau^{8/3}} + \frac{f_{10}(\trho;\te_0,\eta_0)}{\ttau^{10/3}} \, , \label{totalL1}
\end{eqnarray}
where we neglect terms with $i > 11$ since they are beyond the validity
regime of our approximation of the boost-invariant metric.\footnote{In
principle there may be a finite $f_{12}$, but we will ignore it
since it is a higher order term.}
The numerical plots of the non-trivial
profiles $f_i$ ($i=0,4,8,10$) are shown in Fig.~\ref{Fig.f}.

\begin{figure}[]
    \centering
    \subfigure[\label{Fig.taua}$\ttau = 4$]
    {\includegraphics[width=7cm]{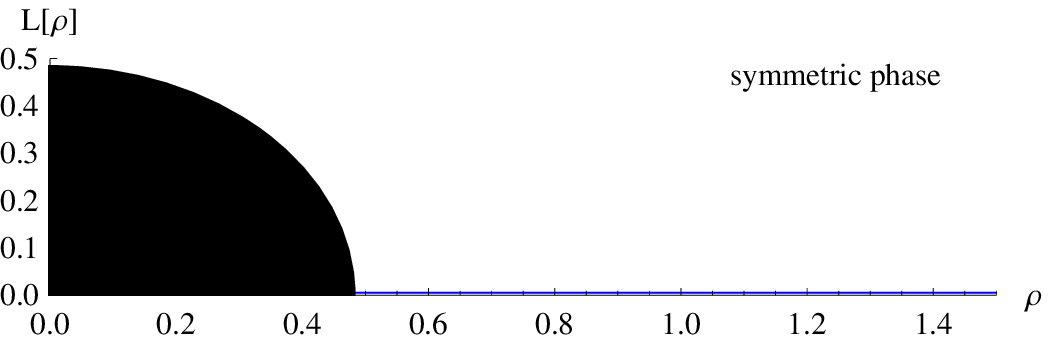}}{\hspace{0.5cm}}
    \subfigure[\label{Fig.taub}$\ttau =14$ ]
    {\includegraphics[width=7cm]{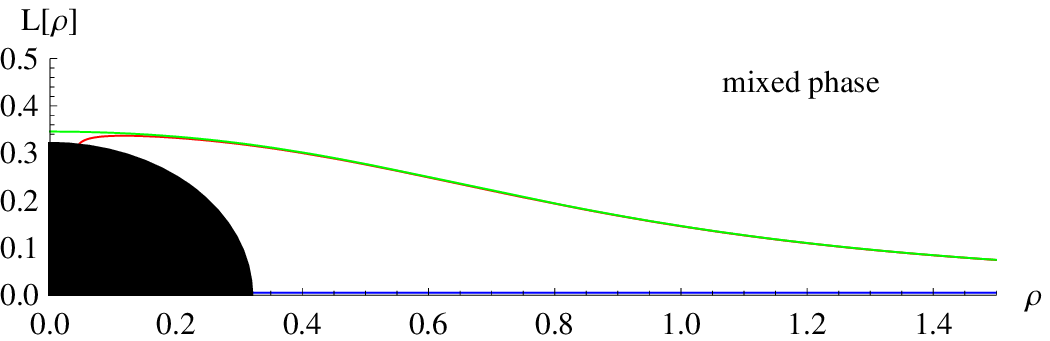}}
    \subfigure[\label{Fig.tauc}$\ttau = 22$]
    {\includegraphics[width=7cm]{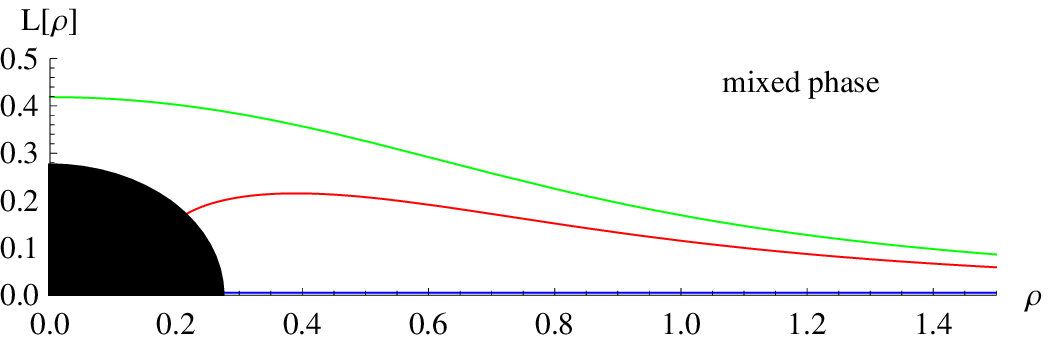}}{\hspace{0.5cm}}
    \subfigure[\label{Fig.taud}$\ttau = 43$]
    {\includegraphics[width=7cm]{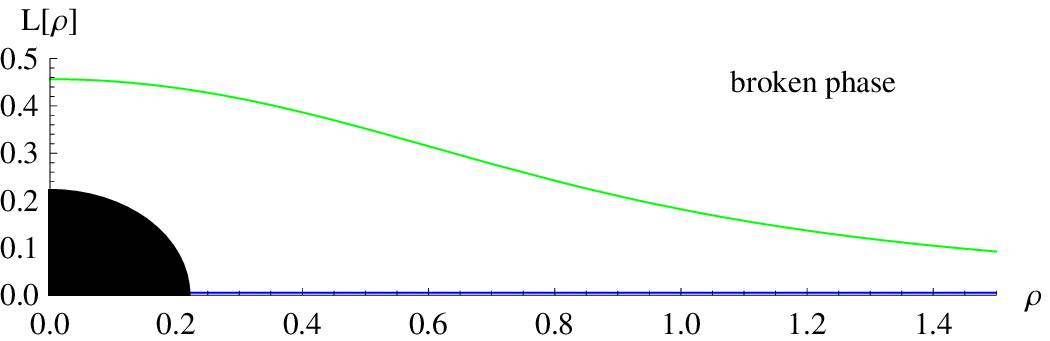}}
        \caption{\label{Fig.tau} Embedding profiles $L(\ttau, \trho)$ obtained from the adiabatic ODE expansion method
         for $\te_0 = 1, \eta_0 =0.3102$
        before (a), during (b,c) and after the phase transition (d): Minkowski embedding (green),
        black hole embedding (red), flat embeddings (blue). }
\end{figure}

These profiles have the following qualitative properties.  As
compared to the solution (\ref{L1}) for  $\nB = 0$, there are
additional non-trivial profiles, $f_0$ and $f_4$.  The profile
$f_0$ agrees with that in the static case ($\ttau \ra
\infty$)~\cite{Johnson1} (dashed curve in Fig.~\ref{Fig.f}a)
We found that the equation for $f_0$ is independent of $\te_0$ and
$\eta_0$, which is natural since at $\ttau \ra \infty$ the system
achieves equilibrium at low temperature and will not depend on the
non-equilibrium dynamics ($\eta_0$) or specific initial conditions
($\te_0$) anymore. The profiles $f_4$ and $f_8$ depend on $\te_0$
and have a negative  sign, which reflects the attraction of the D7
brane to the black hole. $f_{10}$ is the first profile which
depends on $\eta_0$. In contrast to $f_4$ and $f_8$, $f_{10}$ has
a positive sign showing the dragging effect, as in the zero $B$
case around (\ref{c1}). The big amplitudes of $f_4$, $f_8$, and
$f_{10}$ will be suppressed at large $\ttau$ by negative powers of
$\ttau$. For self-consistency we will only consider the $\ttau$
region where all the sub-leading terms are well dominated by the
leading terms, \ie $f_0 \gg f_4 \ttau^{-3/4} \gg \cdots $. Some of
the final embedding profiles $\tL(\ttau,\trho)$ are shown in
Fig.~\ref{Fig.tau}, where the green lines are plotted by plugging
the numerical data of Fig.~\ref{Fig.f} into (\ref{totalL1}).

We would not expect this expansion approach to work beyond the point
where the symmetry breaking embedding ceases to be even a local minimum
because higher order terms will grow. One can nevertheless plot solutions using
$f_0-f_{10}$ that reach down all the way to the horizon. We will use the full
PDE solutions in the next section to test the point where the expansion has
broken down.

In fact early in our studies we tried to use the expansion even for black hole embeddings.
The ODEs for the $f_i$ do not contain the horizon however. We attempted to put the horizon in by hand
by imposing boundary conditions relevant to a black hole. For each point in the
$L-\rho$ plane if we assume a horizon is present we can deduce the value of $\tau$ from
(\ref{rtau}) - one can impose some boundary condition such as orthogonality to the posited
horizon and seek solutions for each $f_i$. This correctly gives one massless solution at each
$\ttau$ and they look very similar to the equilibrium black hole embeddings. In fact though on solving
the full PDEs we realize that this is far beyond the point where the expansion method has collapsed!
We will include some of the resulting curves below though as evidence of the break down of the
ODE approximation.

Fig.~\ref{Fig.tau}
shows the embedding profiles at various stages of the evolution of the
expanding plasma.
The black hole horizon (\ref{rtau}) is indicated by a black quarter circle in the figure.
Its size decreases with time corresponding to a cooling and expanding
plasma. Fig.~\ref{Fig.tau} reflects the division of the
quark-gluon plasma into the three phases described above.

The quark condensate as a function of time can be read form the asymptotic form of the solutions
\begin{eqnarray}
  \tL(\ttau,\trho) \sim \frac{\tc(\ttau)}{\trho^2} \ .
\end{eqnarray}
We plot these results in Fig.~\ref{Fig.c}a (the dashed lines) for comparison to the
equilibrium inspired results. The main result here is that the late time behaviour is
indeed just the equilibrium expectations. Where the curved line deviates from the
equilibrium results is in fact a sign that the expansion used in this section has
broken down, the configuration is out of equilibrium and adiabatic behaviour is no longer
possible. Full solutions of the PDEs will show this in the next section.

\subsection{Full PDE solutions}

In the previous section we interpreted our solution obtained from
the ordinary differential system as the evolution of extremum
states in the potential. This evolution also reflects the real
time dynamics of the plasma whenever there is a unique deep global
minimum for the vacuum state, because the embedding time dynamics
is expected to be well localized around the minimum. This is the
case in the chiral symmetric phase at early times and in the
broken phase at late times. In the mixed phase at intermediate
times there appears also a local minimum in addition to the global
minimum. The small potential barrier separating these minima is
potentially easily overcome by a fluctuation.
The path of the local maximum seen in the adiabatic approach will
not be realized as a dynamical solution  since it is
unstable. Further below the critical time $\ttau_a$ where the
symmetry breaking minimum disappears a heating vacuum will be left
in a very non-equilibrium state that again can not be followed
using the expansion technique of section~\ref{secadiabatic}.

To find the real time evolution of the chiral transition out of
equilibrium, we need to solve the PDE directly and compare the
solution with our previous approximate solutions. In practice it
is more convenient to consider a heating process than a cooling
one\footnote{{In principle, time reversed heating is justified only at zero viscosity since 
a finite viscosity results in decreasing entropy. However, 
we keep a finite viscosity in our numerics since it has a negligible effect and 
the results are relevant to the cooling case.    
}} because we can use a well defined starting configuration at
large $\ttau$ as an initial condition of our partial differential
equation, \ie
\begin{eqnarray}
  \tL(\ttau \ra \infty, \trho) = f_0 \ , \qquad \del_\ttau\tL(\ttau \ra \infty, \trho) = 0 \ ,\label{BC1}
\end{eqnarray}
with $f_0$ as in~(\ref{totalL1}).
For simplicity, we impose only Neumann boundary conditions at $\trho=0$ associated to
Minkowski embeddings and a zero bare quark mass condition at $\trho \ra \infty$ to study
the spontaneous symmetry breaking,
\begin{eqnarray}
  \del_\trho \tL(\ttau,\trho = 0)=0 \ , \qquad  \tL(\ttau,\trho \ra \infty)=0
  \, . \label{BC2}
\end{eqnarray}
The conditions (\ref{BC1}) and (\ref{BC2}) completely determine the dynamics with
the partial differential equation derived by varying the action (\ref{action2}).
Again, the actual expression of the equation of motion is lengthy and will not be shown here.

Numerically we solve the equation using Mathematica's inbuilt PDE
solvers. These are somewhat temperamental and one needs to spend
considerable time adjusting precision tolerances in order to find
smooth solutions in sensible periods of computer time. When we
have such solutions we test their stability to changes in
precision settings to ensure they are reliable.

With the boundary conditions above we can run our simulations
until the embeddings touch the black hole. Beyond that one needs dynamic
boundary conditions along the black hole surface. At least in the
coordinates we use here this is a hard problem. We have found a
simple trick that seems to produce sensible black hole embeddings
though. After the D7 has touched the horizon at $\trho=0$ we
artificially hold the embedding at the top of the horizon. The
large $\trho$ evolution of the D7 is local and relatively
unaffected by this incorrect embedding at small $\trho$. 
\begin{figure}[]
    \centering
    {\includegraphics[width=11.5cm]{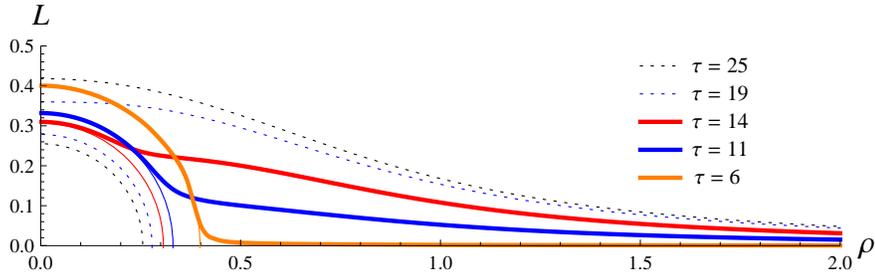}} \hspace{1cm}
\caption{\label{superheat} The embedding profiles $\tilde L(\ttau, \trho)$ 
from the full PDE solution for a solution
starting in the low temperature symmetry breaking vacuum
($\kappa=1$). }
\end{figure}
\begin{figure}[]
    \centering
    {\includegraphics[width=8cm]{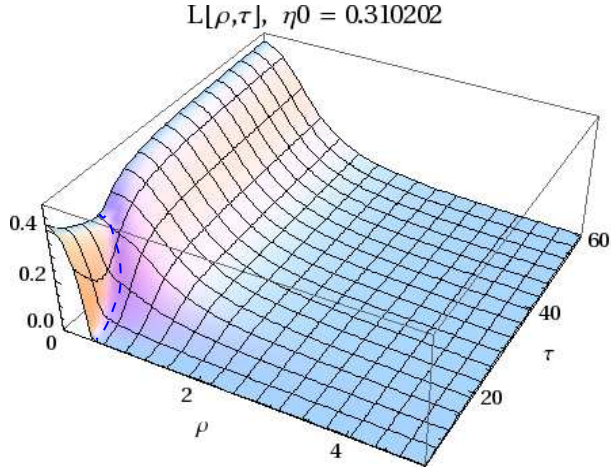}} \hspace{1cm}
\caption{\label{3dplot} 3D plot of the embedding profile $\tilde L(\ttau, \trho)$.  }
\end{figure}
Further,
as has been observed before, solutions shooting from the black
hole horizon experience a numerical attraction onto the unique
regular black hole embedding ending at a given point on the
horizon. The result is that we get numerical solutions like those
shown in Fig.~\ref{superheat} (\eg the red, blue, or orange curve)
 where the D7 follows the horizon before
shooting out to large $\trho$. We believe that these represent very
good approximations to the large $\trho$ embeddings solutions.
Since we extract the condensate $\tc$ at large $\trho$ we will live
with the improper near horizon behaviour. It would of course be
interesting to try to improve on this with dynamic boundary
conditions in the future.

As a first example of a solution we will study the super-heated
symmetry breaking vacuum. At large $\ttau$ we use the leading
terms in the expansion from Section 3.2 to find the UV
configuration - one needs to use several terms in the expansion to
find the embedding with no extra energy. We then solve for the
evolution to low $\ttau$. In Fig.~\ref{superheat} we show plots of
the embedding $\tilde L(\trho)$ for different $\ttau$ with the black hole's
position for each $\ttau$ also shown. We expect that the near horizon behaviour 
is not correct but the far UV embedding should approximate 
the solution we seek well. In the 3d plot of $\tL(\ttau,\trho)$
 shown in Fig.~\ref{3dplot} this corresponds to excising the interior of the region
 indicated by the dashed blue line.
 
A smooth evolution is apparent. To compare this to our various
approximations above we also plot the condensate as a function of
time in Fig.~\ref{Fig.c}a (blue curve). Again the solution follows
the equilibrium estimate and the expansion solution at large
$\ttau$. In the period $\ttau_{a-c}$ it follows the equilibrium
result not the ODE expansion results showing that expansion had
broken down before the brane  left the mixed phase. The success of
the equilibrium approximation suggests we should take its estimate
of the transition points $\ttau_a$ (where the local symmetry
breaking minimum vanishes) and $\ttau_c$ (where the two minima of
the mixed phase are degenerate in energy) as correct. The full
PDE solutions allow us to know in addition the behaviour of the
condensate when we have heated above the temperature where the
super-heated phase has stopped having a local minima (beyond
$\ttau_a$). This is the main result of our analysis here.

\begin{figure}[]
    \centering
    {\includegraphics[width=8cm]{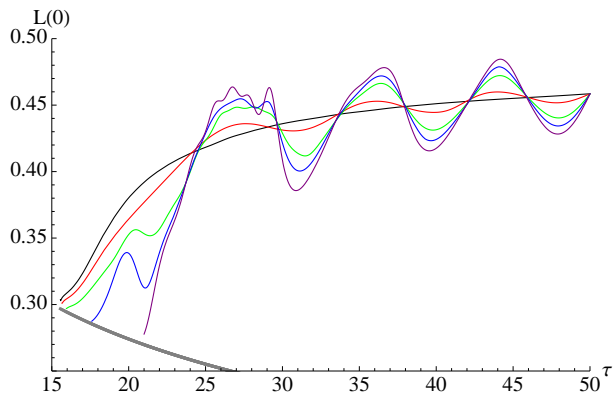}} \hspace{1cm}
\caption{\label{excite} Plots of the IR position of the D7 brane
against time for a number of large time initial conditions with
different energy. Initial velocities: black ($\kappa=1$), red
($\kappa=10$), green ($\kappa=30$), blue ($\kappa=40$), purple
($\kappa=50$). The thick black line is the horizon.}
\end{figure}

It would be nice to test the equilibrium configurations estimate
for the length of time in which the mixed phase exists
($\ttau_b-\ttau_a$ which is about 10 units of $\ttau$). This we
can do by looking at some simple out of equilibrium
configurations. For the plots so far shown we computed
$\dot{\tL}(\trho)$ at some large $\ttau$ from the $f_4$ term in the
expansion for $\tL(\trho, \ttau)$. We can give the configuration more
energy by simply multiplying that $\dot{\tL}(\trho)$ by a numerical
factor, $\kappa$~\footnote{
In practice, we fix some $\ttau_{\mathrm{max}}$ and modify the initial
conditions (\ref{BC1}) as $L(\ttau \rightarrow \ttau_{\mathrm{max}}, \trho) = L^{ODE} (\ttau_{\mathrm{max}}, \trho)$ and $\dot L(\ttau \rightarrow \ttau_{\mathrm{max}}, \trho) = \kappa * \dot
L^{ODE} (\ttau_\mathrm{max}, \trho)$, where $L^{ODE}$ is given by (\ref{totalL1}).}. 
One might think of these initial states as
thermally excited versions of the asymptotic vacuum. In
Fig.~\ref{excite} we plot the evolution of $\tL(\trho = 0)$ as a function
of $\ttau$ for a number of such states. At large $\ttau$ when the
theory is cold the unique vacuum is the symmetry breaking one and
with extra energy the configuration oscillates about that minimum.
The motion is simple harmonic as can be seen from the independence
of the period on the amplitude of the oscillations. As the
solutions approach $\ttau_c$ the different solutions begin to
diverge. The solution with $\kappa=10$ lies close to the
equilibrium curve - the oscillations about the minimum are small
and the state stays super-heated. When $\kappa=30,40$ $\tL(0)$ falls
more quickly suggesting that at least some of the brane's length
in $\rho$ has escaped the local potential minimum. The upwards
wiggles suggest that some of the length is still repelled back
into the well though. Finally though by $\kappa=50$ the brane has
ridden over the potential barrier and escaped the local minimum.
The difference in arrival times at the horizon for these
configurations (about 6 units of $\ttau$) is a rough estimate of
the period of the mixed phase. It seems to broadly match the
equilibrium inspired picture again.
\begin{figure}[]
    \centering
    {\includegraphics[width=8cm]{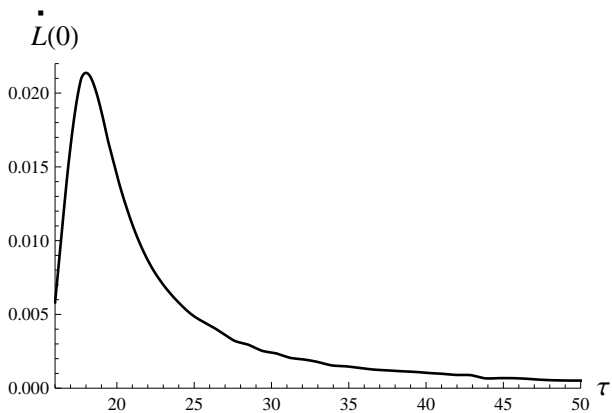}} \hspace{1cm}
\caption{\label{velocity} The IR speed of the D7 brane
against time for the $\kappa=1$ initial condition. }
\end{figure}

In Fig.~\ref{velocity} we plot the motion of the IR $\rho=0$
end point of the brane with time for a low energy configuration
($\kappa \simeq 1$). The brane certainly seems localized in a
minimum down to a time of order $\ttau=27$ (compare to the
equilibrium estimate that the mixed phase begins at $\ttau_b=34.9$
and becomes meta-stable at $\ttau_c=27.5$). Further the steepest
period of acceleration is below $\ttau=22$ (to be compared with
the equilibrium estimate for when the metastable vacuum ceases to
exist $\ttau_a=20.6$). We conclude that this system is indeed a
super-heated state that survives in the local minimum until very
close to the equilibrium estimate for~$\ttau_a$. Note the other
obvious feature in the plot is the deceleration just below
$\ttau=20$. This corresponds to where in our simulation the D7
first impacts on the black hole - at smaller $\ttau$ we hold the
D7 at the horizon as discussed above so this behaviour is an
artefact. A full solution would continue to accelerate along the
horizon.

These configurations with excess $\dot{L}$ are also very much
linked to bubble formation. A bubble forms in the mixed phase when
a volume of space has excess energy due to a thermal fluctuation
and escapes the local minima early. Here by treating the whole
space as one we are essentially describing the formation of a
large homogeneous bubble. It would be interesting in the future to
try to study $x$ dependent initial conditions to understand how
quickly or slowly bubbles grow.

\section{Dependence of the condensate on $\nB$}

In the previous section we focused on the chiral transition
induced by a magnetic field, which simply played the
role of an intrinsic symmetry breaking scale.  In this section
we turn to the physics depending on $B$ more quantitatively.

To make analytic progress we will concentrate on the
adiabatic (ODE) approach in which we rescaled all variables by
some power of $\nB$, see (\ref{Dimless}).  In order to study the
dependence of the quark condensate on $\nB$, we use the original
parameters, in terms of which the condensate can be expanded as
\begin{eqnarray}
  && c(\tau, \nB) =  \nB^{3/2} \left(c_0
  + \frac{1}{\nB^{2/3}}\frac{c_4(\nB^{-4/3}\ve_0)}{\tau^{4/3}}
  + \frac{1}{\nB^{4/3}}\frac{c_8(\nB^{-4/3}\ve_0)}{\tau^{8/3}} \right. \nn \\
  && \quad \qquad \qquad \qquad \left.
  + \frac{1}{\nB^{5/3}}\frac{c_{10}(\nB^{-4/3}\ve_0,\eta_0)}{\tau^{10/3}} \right)\,.
\end{eqnarray}
Note that ${c}_0$ is independent of $\ve_0, \eta_0$ and $\eta_0$
enters only in ${c}_{10}$.  The $\nB$ dependence of the leading
term agrees with that in the static (zero temperature) case~\cite{Johnson1}, {\em i.e.}\ it
scales as $\nB^{3/2}$.  The first subleading term ($\sim \tau^{-4/3}$)
may be compared to the finite temperature case~\cite{Johnson1,
Evans1}. In the adiabatic approximation, where $T \sim \tau^{-1/3}$,
this term scales approximately like $T^4 \nB^{-5/6}$. However, due to
the $B$ dependence of $c_4$, this scaling is not exact. In general,
the effect of the subleading terms is to lower the exponent in
$c(\tau,\nB) \sim \nB^\nu$ to a value $\nu < 3/2$.
The scaling of the total condensate $c(\tau, \nB)$ with $\nB$ will again
be determined numerically.
\begin{figure}[]
    \centering
    \subfigure[\label{Fig.Ha}]
    {\includegraphics[height=6.1cm]{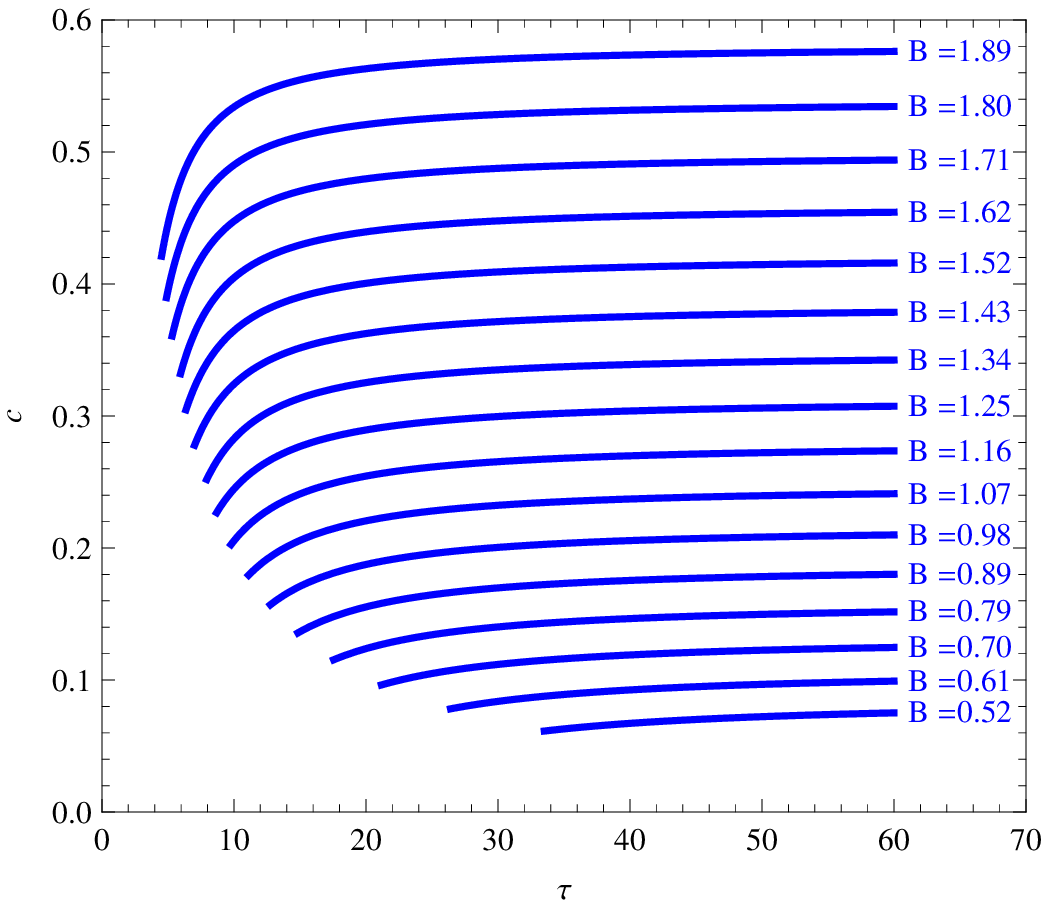}} \hspace{1cm}
    \subfigure[\label{Fig.Hb}]
    {\includegraphics[width=6cm]{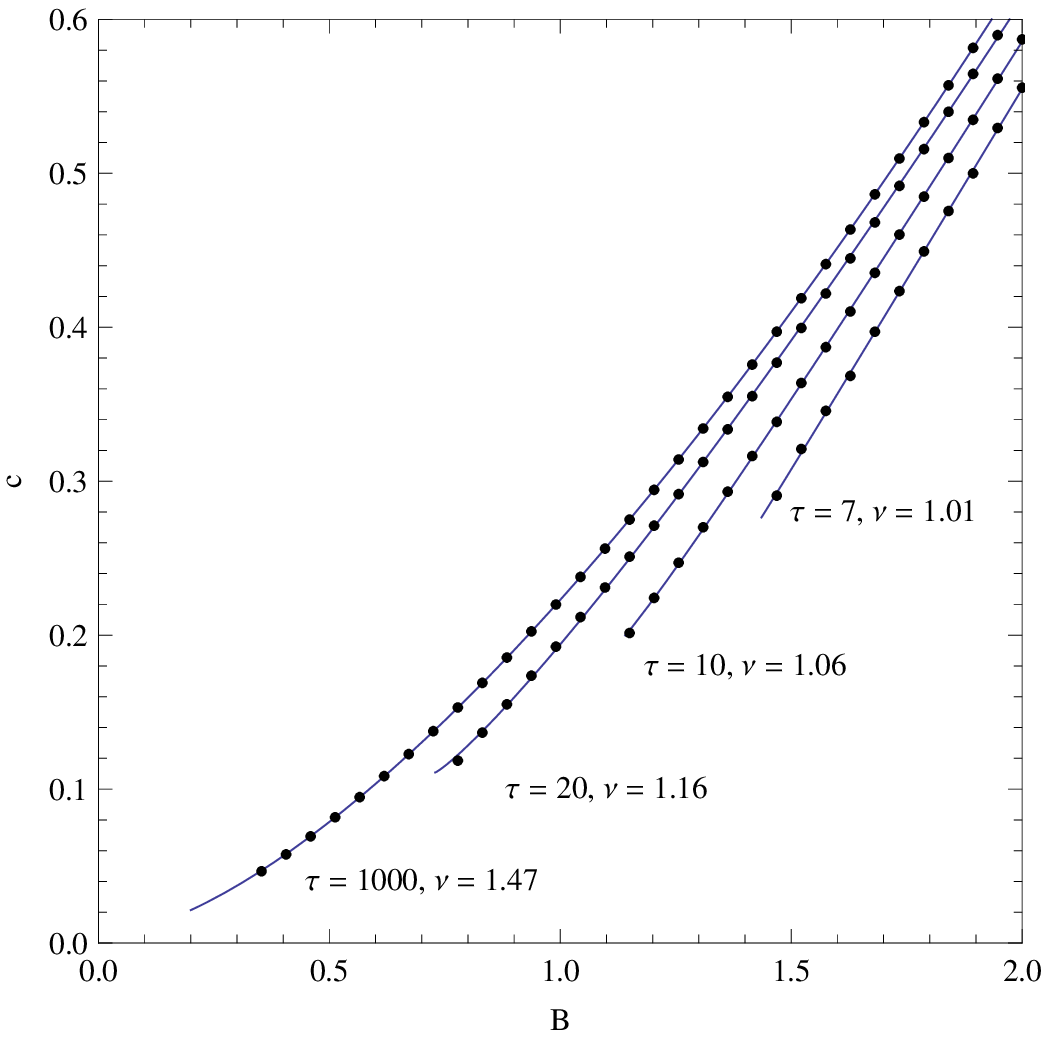}}
\caption{\label{Fig.H} The condensate $c(\tau, \nB)$ $(\ve_0=1)$
from the ODE adiabatic approximation results. (a) condensate
$c(\tau)$ for various $\nB$. (b) showing $c(\nB)\propto \nB^\nu$
for $\tau=7,10,20,1000$] }
\end{figure}

Fig.~\ref{Fig.H}a shows our results for the condensate for various
values of the magnetic field. We have only plotted the condensates
associated with the Minkowski embeddings, which exist for some
time $\tau > \tau_*(B)$. Here $\tau_*$ is defined as the time,
when the Minkowski solution meets the black hole solution (This
corresponds to the cusp in the dashed curve in Fig.~\ref{Fig.c}a).
We take $\tau_*$ to mark the time of the phase transition. As
can be seen from Fig.~\ref{Fig.c}a the early time behaviour is an
imperfect approximation to the full PDE solutions but the later
time solutions are consistent with the solutions. We find that
for fixed $\tau$, or equivalently for fixed temperature, the
condensate grows with increasing magnetic field. In the limit
$\tau\ra\infty$ ({\em i.e.}\ at zero temperature), we find the
following dependence on $\nB$:
\begin{equation}
\lim\limits_{\tau\ra\infty} c(\tau) = c_0 \nB^{3/2} = 0.223 \nB^{3/2},
\end{equation}
which is in agreement with the zero temperature result \cite{Johnson1}.
For earlier times, the dependence on $\nB$ is shown in
Fig.~\ref{Fig.H}b.  We find numerically that $c(\tau, \nB) \sim
\nB^\nu$ where the power $\nu$ decreases from $1.5$ at large
$\tau$ to approximately $1.0$ at small $\tau$. In other words, the
dependence on $\nB$ tends to become linear at high temperatures.
Our results hold for sufficiently strong magnetic fields. At small
$B$ the system is in the symmetric phase ($c=0$).

\begin{figure}[]
    \centering
    {\includegraphics[width=6cm]{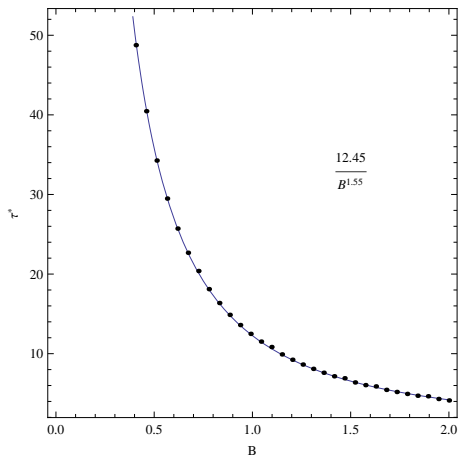}}
\caption{\label{Fig.G} $\tau_*$ as a function of $\nB$ $(\ve_0=1)$
from the ODE adiabatic approximation results. }
\end{figure}

The tendency for the condensate to increase with $B$ is in
qualitative agreement  with
observations in chiral perturbation theory \cite{Schramm,
Shushpanov} ($\propto B^{3/2}$ for strong fields,  $\propto B$ for
weak fields), in the Nambu-Jona-Lasinio model \cite{Klevansky}
($\propto B^2$), in a confining deformation of the holographic
Karch-Katz model \cite{Zayakin} ($\propto B^2$), and in SU(2)
\cite{Buividovich} ($\propto B$) and SU(3) \cite{Kalaydzhyan}
lattice calculations ($\propto B^{1.6\pm 0.2}$). The
dependence on $\nB$ typically ranges
from linear to quadratic behaviour, {\em i.e.}\ the powers of
$\nB$ are in the range $1 \leq \nu \leq  2$.

We may also study the effect of $\nB$ on the time of the chiral
transition which is marked by~$\tau_*$.  From the D-brane picture
we expect that $\tau_*$ decreases with $\nB$.  For large $\nB$,
the repelling force caused by the B-field is much stronger than
the attractive force of the black hole. Even at early times,
Minkowski embeddings, associated to stable mesons, are therefore
favoured over black hole embeddings, which implies that the meson
melting process sets in at some earlier time, {\em i.e.}\ at
higher temperatures.

Fig.~\ref{Fig.G} shows $\tau_*$ as a function of $\nB$.  We find that
$\tau_*$ indeed decreases with increasing~$\nB$ as
$\tau_*(\nB)\sim\nB^{-1.55}$.  The numerically found exponent $-1.55$
for the scaling of $\nB$ is close to $-1.5$ for the case of vanishing
shear viscosity, which can be
explained as follows. At $\ttau = \ttau_*$ and $\eta=0$,
the horizon is located at
\begin{equation}
 \tr_* = \frac{\te_0^{1/4}}{3^{1/4} \ttau_*^{1/3}} =
 \frac{ \ve_0^{1/4}}{3^{1/4}\sqrt{\nB}} \t^{-1/3}_* = \frac{1}{3^{1/4}(12.45)^{1/3}} \ ,
\end{equation}
where the last equality is from the
numerical value of $\tr_*$ at $\te_0=1$.  Thus
\begin{equation}
\t_* = 12.45 \nB^{-1.5}\,,
\end{equation}
where for the numerical analysis we chose $ \ve_0 = 1$. The deviation
from $-1.5$ is due to the effect of the shear viscosity.
We also numerically confirmed that the exponent is $1.5$ without viscosity.

This scaling of $\tau_*$ may be compared to results for the critical
temperature $T_c$ of the phase transition in the static case. In the
adiabatic approximation (when $\eta_0=0$), $T \sim \tau^{-1/3}$, which
implies $T_* \sim \nB^{1/2}$.  This square root behaviour is in
agreement with the result for the critical temperature $T_c$
in the static approach~\cite{Evans1} and with \cite{Shushpanov} for strong
magnetic fields. It is also in qualitative agreement with
studies in QCD \cite{Chernodub, Elia}.

\begin{figure}[]
    \centering
    {\includegraphics[width=6cm]{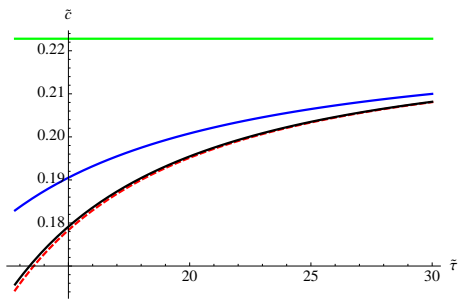}}
\caption{\label{conds} The condensate $\wt{c}(\ttau)$. The green curve is the leading term, the blue 
is up to subleading term, the red is up to third term and the black is the whole expression. }
\end{figure}

Finally we consider the effect of changing the viscosity. 
The viscosity effect is very small, since it is doubly
suppressed by both large $\ttau$ and small $\eta_0$. This can also
be seen in Fig.~\ref{conds}, where the condensate
 is plotted for four cases: the green is the
leading term, the blue includes the subleading term, the red
includes up to the third term and the black is the full
expression. The viscosity effect is then the difference between
the red and black curves. There is a small visible difference only
at small $\ttau$.

\section{Discussion}

We have analyzed the first order chiral and meson melting phase
transition in a warming or cooling strongly coupled gauge theory
with quarks using the AdS/CFT Correspondence. We have developed
numerical techniques to study the PDE that describes the motion of
a D7 brane in a time dependent geometry. In particular this allows
us to explicitly find smooth solutions of the non-equilibrium
configurations that are necessarily part of the transition. These
results confirm the equilibrium analysis of the transition but
also go beyond them. For example we have described the formation
of large homogeneous bubbles in the mixed phase of the transition.
In the future it would be interesting to study spatially
inhomogeneous bubbles to understand their growth although this
would require the solution of a 2+1 dimensional PDE which is
potentially more numerically tricky.

To keep our problem a simpler 1+1 dimensional PDE we also
restricted motion of the D7 brane to the holographic coordinates
$L$ and $\rho$. There could also of course be fluctuations in the
holographic angular direction $\phi$ which we have not described.
Such configurations might be useful in the study, for example, of
disordered chiral condensates \cite{Rajagopal:1992qz}.
We may also consider the finite density case 
by turning on the time component of the
$U(1)$ gauge field on the probe brane~\cite{Evans1,Kim2}.

We hope that the system we have studied can shed some light on
first order transitions in a range of strongly coupled gauge
theories including perhaps QCD. The ${\cal N}=4$ background,
however, does deconfine in the presence of an infinitesimal
temperature, which is not expected in simple QCD. We
note though that there are several arguments in favour of a possible
splitting of the deconfinement and chiral transitions in QCD in
the presence of a strong magnetic field~\cite{Chernodub}.
While the temperature of the chiral transition increases~\cite{Fraga, Chernodub},
the temperature of the deconfining transition either decreases with
increasing magnetic field~\cite{Agasian} or increases but much slower than 
the chiral transition \cite{Chernodub}. Both transitions
become first-order transitions and a new phase with broken
chiral symmetry and deconfinement appears for sufficiently strong magnetic
fields. So, it is the region of the $T$-$\nB$ phase diagram
with strong magnetic fields and above the deconfinement temperature
in which our model might be qualitatively compared with QCD.

\acknowledgments

The authors thank Romuald Janik, Johanna Erdmenger,  Jonathan Shock, Michal Heller, 
Volker Schomerus and Piotr Surowka for discussions.
NE and KK are grateful for support from STFC.

\end{document}